\documentclass[aps,prb,showpacs,floatfix,twocolumn,byrevtex,superscriptaddress]{revtex4-1}
\usepackage{amsmath}
\usepackage{amssymb}
\usepackage{amstext}
\usepackage{amsopn}
\usepackage{amsfonts}
\usepackage{amsxtra}
\usepackage[english]{babel}
\usepackage{amsmath}
\usepackage{graphicx}
\usepackage{float}
\usepackage{bm}
\usepackage{multirow}
\usepackage{dcolumn}
\usepackage{hyperref}
\begin{document}

\title{Spin-flop transition and magnetic phase diagram in CaCo$_{2}$As$_{2}$ revealed by torque measurements}

\author{W. Zhang}
\affiliation{College of Physics, Optoelectronics and Energy $\&$ Collaborative Innovation Center of Suzhou Nano Science and Technology, Soochow University, Suzhou 215006, China}
\affiliation{Beijing National Laboratory for Condensed Matter Physics, National Laboratory for Superconductivity, Institute of Physics, Chinese Academy of Sciences, P.O. Box 603, Beijing 100190, China}

\author{K. Nadeem}
\email[]{kashif.nadeem@iiu.edu.pk}
\affiliation{Beijing National Laboratory for Condensed Matter Physics, National Laboratory for Superconductivity, Institute of Physics, Chinese Academy of Sciences, P.O. Box 603, Beijing 100190, China}
\affiliation{Department of Physics, International Islamic University, Islamabad, Pakistan}

\author{H. Xiao}
\email[]{hxiao@iphy.ac.cn}
\author{R. Yang}
\author{B. Xu}
\affiliation{Beijing National Laboratory for Condensed Matter Physics, National Laboratory for Superconductivity, Institute of Physics, Chinese Academy of Sciences, P.O. Box 603, Beijing 100190, China}

\author{H. Yang}
\email[]{yanghao@nuaa.edu.cn}
\affiliation{College of Physics, Optoelectronics and Energy $\&$ Collaborative Innovation Center of Suzhou Nano Science and Technology, Soochow University, Suzhou 215006, China}
\affiliation{College of Science, Nanjing University of Aeronautics and Astronautics, Nanjing 211106, China}

\author{X. G. Qiu}
\affiliation{Beijing National Laboratory for Condensed Matter Physics, National Laboratory for Superconductivity, Institute of Physics, Chinese Academy of Sciences, P.O. Box 603, Beijing 100190, China}

%
%
%

\begin{abstract}

The magnetic properties of CaCo$_{2}$As$_{2}$ single crystal was systematically studied by using dc magnetization and magnetic torque measurements. A paramagnetic to antiferromagnetic transition occurs at $T_N$ = 74 K with Co spins being aligned parallel to the c axis. For $H \parallel c$, a field-induced spin-flop transition was observed below $T_N$ and a magnetic transition from antiferromagnetic to paramagnetic was inferred from the detailed analysis of magnetization and magnetic torque. Finally, we summarize the magnetic phase diagram of CaCo$_{2}$As$_{2}$ based on our results in the \emph{H-T} plane.

\end{abstract}



\maketitle

%

\section{INTRODUCTION}

In an uniaxial low anisotropic antiferromagnet, a sufficiently strong magnetic field applied along the easy axis could induce a spin flop (SF) transition.\cite{Neel1952} This response to the magnetic field has been observed in many antiferromagnetic (AFM) materials, such as AFM semiconductors,\cite{Fries1997} organic magnets,\cite{Uozaki2000,Sasaki2001,Wernsdorfer2002} organic superconductors,\cite{Kawamoto2008} and nanomagnetic systems.\cite{Fitzsimmons2000} In the last decades, several magnetic studies in the materials with a layered ThCr$_{2}$Si$_{2}$ type(122) crystal structure have been reported,\cite{Reehuis1990,Reehuis1993,Reehuis1994,Reehuis1998} which again come to fore in recent researches after the discovery of superconductivity in 122 iron-based family.\cite{Rotter2008,Rotter2008a} The parent compounds of these iron-based superconductors (FeSCs) are found to be AFM. The AFM order can be suppressed via carrier doping (hole or electron) or by applying pressure, and superconductivity emerges in the vicinity of vanishing AFM order. The most striking feature of FeSCs is the coexistence of magnetism and superconductivity,\cite{Cruz2008, Fernandes2010, Shermadini2011} which also differentiates them from cuprates. Analysis of the magnetic transitions and order in the parent compounds can contribute to the understanding of the pairing mechanism in high-\emph{T$_c$} superconductors.\cite{Mazin2008, Dong2010}

CaFe$_{2}$As$_{2}$ parent compound is one interesting material, which shows a nonmagnetic collapsed tetragonal (cT) phase and pressure-induced superconductivity .\cite{Torikachvili2008,Kreyssig2008,Goldman2009,Saha2012} And it exhibits a combined structural and magnetic transition at 170 K.\cite{Ni2008} Recently, it is reported that  CaCo$_{2}$As$_{2}$, the complete replacement of Fe in CaFe$_{2}$As$_{2}$ by Co, exhibits remarkable different properties as compared to CaFe$_{2}$As$_{2}$.\cite{Cheng2012, Ying2012} CaCo$_{2}$As$_{2}$ showed an A-type AFM order with the Co spins oriented ferromagnetically within the \emph{ab} plane and antiferromagnetically along the \emph{c} axis. The AFM aligned Co spins can be flopped by applying a magnetic field along the \emph{c}-axis and eventually aligned along the direction of applied field at a sufficiently strong magnetic field. Cheng et al. reported two successive SF transitions in CaCo$_{2}$As$_{2}$ with a N¡äeel temperature $T_{N}$ = 76 K .\cite{Cheng2012} While Ying et al. reported single SF transition by using magnetization and electronic transport measurements,\cite{Ying2012} they also found that $T_{N}$ is enhanced from 70 K to 90 K and the SF field ($H_{SF}$) is suppressed with 10\% Sr substitution on Ca sites (Ca$_{0.9}$Sr$_{0.1}$Co$_{2}$As$_{2}$). Quirinale et al. and Anand et al. studied the crystalline and magnetic structure of CaCo$_{2}$As$_{2}$ by using X-ray diffraction and neutron diffraction analysis, respectively.\cite{Quirinale2013,Anand2014} The results showed an A-type AFM magnetic structure, which is in accord with that reported by Cheng et al. \cite{Cheng2012} and Ying et al..\cite{Ying2012} Besides, they found that the actual composition of CaCo$_{2}$As$_{2}$ is CaCo$_{1.86}$As$_{2}$ due to 7\% vacancies on the Co lattice sites. The discrepancy in the reports motivate us to make further research about the magnetic transition in this interesting compound CaCo$_{2}$As$_{2}$ by using a different technique, torque magnetometry, which is very sensitive to magnetic transitions.

In this study, we use dc magnetization and magnetic torque measurements to clarify the SF transition in CaCo$_{2}$As$_{2}$ in more detail. We have provided a direct comparison of magnetization measurements with torque magnetometry. Torque magnetometry measures the magnetic torque $\tau= \vec M \times \vec B$ experienced by a sample of magnetic moment $\vec M$ in an applied magnetic field $\vec B$. Compared to longitudinal magnetometry, torque magnetometry is more sensitive to detect the metamagnetism because the magnitude and direction of the magnetization can be extracted simultaneously.\cite{Canfield1997,Naugle2008} Our results in CaCo$_{2}$As$_{2}$ single crystal reveal an A-type AFM structure, a field-induced SF and magnetic transition from AFM to PM state below \emph{T$_{N}$}. Finally, we give a detailed magnetic phase diagram in the \emph{H-T} plane of CaCo$_{2}$As$_{2}$ single crystal.

%
%

\section{EXPERIMENTAL DETAILS}

%
\begin{figure}[tb]
\includegraphics[width=0.9\columnwidth]{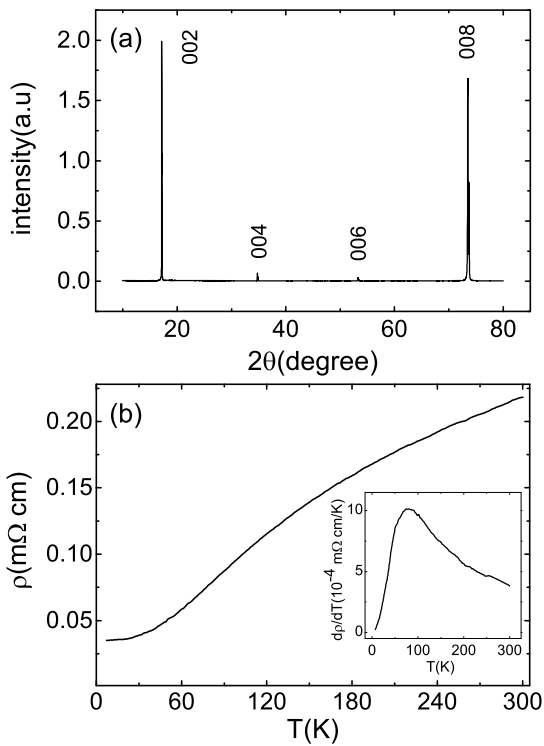}
\caption{(color online) (a) The typical XRD $\theta$-2$\theta$ scan of CaCo$_{2}$As$_{2}$ single crystal. (b) $\rho$ vs \emph{T} curve with \emph{I} // \emph{ab} at \emph{H} = 0 T. Inset: $d\rho/dT$ vs \emph{T} curve.}
\label{Fig1}
\end{figure}

The CaCo$_{2}$As$_{2}$ single crystals were grown by using CoAs self-flux method as described elsewhere.\cite{Goldman2009,Wang2009}  The crystal structure was investigated by X-ray diffraction (Rigaku D/MAX-Ultima III) using Cu K$_{\alpha}$ ($\lambda$ = 0.154 nm) radiation at ambient conditions. The actual elemental compositions were analysed by inductively coupled plasma-atomic emission spectrometry (ICP-AES, IRIS Intrepid II XDL) and energy dispersive X-ray (EDX) detector using a Hitachi S-4800 scanning tunnel microscope. Magnetization measurements were done by using superconducting quantum interference device (SQUID, Quantum Design) and Physical Property Measurement System (PPMS-9, Quantum Design). The magnetic torque was measured as a function of angle and magnetic field over a wide temperature range by using a piezoresistive magnetometry. In piezoresistive magnetometry, the torque lever chip together with a puck, is mounted on a PPMS horizontal rotator. The sample is mounted on the center of chip by using Apiezon N grease. A wheatstone bridge circuit (integrated on the chip) detects the change in the resistance of the piezoresistors, produced by the different magnetic torque. The samples for all measurements presented here are from the same batch.

%

%
%

\section{RESULTS AND DISCUSSION}

\subsection{Crystal structure and EDX,ICP-AES} 

Fig. 1(a) shows a typical XRD $\theta$-2$\theta$ scan of CaCo$_{2}$As$_{2}$ single crystal. All the indexed peaks are identified as the ThCr$_{2}$Si$_{2}$ body-centered tetragonal structure phase. The presence of only (00\emph{l}) reflection peaks indicates that \emph{c}-axis is perpendicular to the crystal plane. The calculated lattice constant of c-axis comes out to be $\sim$10.3 \AA,  close to the previous report, where $a=b=3.9906(1)$ {\AA}, $c=10.2798(2)$ {\AA}.\cite{Quirinale2013}

We have also done temperature dependent resistivity $\rho(T)$ (\emph{I} $\parallel$ \emph{ab}) measurements in the temperature range from 5 to 300 K (Fig. 1(b)). There is no upturn resistivity associated with the AFM transition observed at \emph{T$_N$} in $\rho$(T) curve, which is different from other 122 parent compounds, such as CaFe$_{2}$As$_{2}$ and BaFe$_{2}$As$_{2}$,\cite{Ronning2008,Rotter2008,Rotter2008a} where clear upturn resistivity were observed below their AFM transition temperature $T_{N}$. As shown in the inset of Fig. 1(b), a broad peak at 74 K is observed in $d\rho/dT$, such a broad peak in $d\rho/dT$ has been associated with the antiferromagnetic transition in Ref.~\onlinecite{Ying2013}.

\begin{table}[tp]
\caption{\label{tab1} Actual composition of CaCo$_{2}$As$_{2}$ single crystal obtained from EDX and ICP-AES.}
\begin{ruledtabular}
 \begin{tabular}{ccccccc}
  &      &   EDX    &          &               &    ICP-AES    &           \\
\cline{2-4}\cline{5-7}
Sample   &  Ca  &   Co     &    As    &       Ca      &        Co     &       As  \\
\hline
\#1 &   1  &    1.93  &    2.01  &      1.02     &       1.90    &       2   \\
\#2 &   1  &    1.89  &    2.08  &        1      &       1.92    &       2   \\
\#3 &   1  &    1.97  &    2.11  &        1      &       1.90    &       2   \\
\end{tabular}
\end{ruledtabular}
\end{table}

Table 1 shows the results of EDX and ICP-AES measurements on three samples of CaCo$_{2}$As$_{2}$. From the EDX data we conclude that the crystal have some vacancies on the Co sites, from the ICP-AES data we estimate about a 5$\%$ vacancies on the Co sites, corresponding to the actual composition CaCo$_{1.9}$As$_{2}$, which is very close to reported composition CaCo$_{1.86}$As$_{2}$ in Ref.~\onlinecite{Quirinale2013}.

\subsection{Magnetization} 

%
\begin{figure}[tb]
\includegraphics[width=0.9\columnwidth]{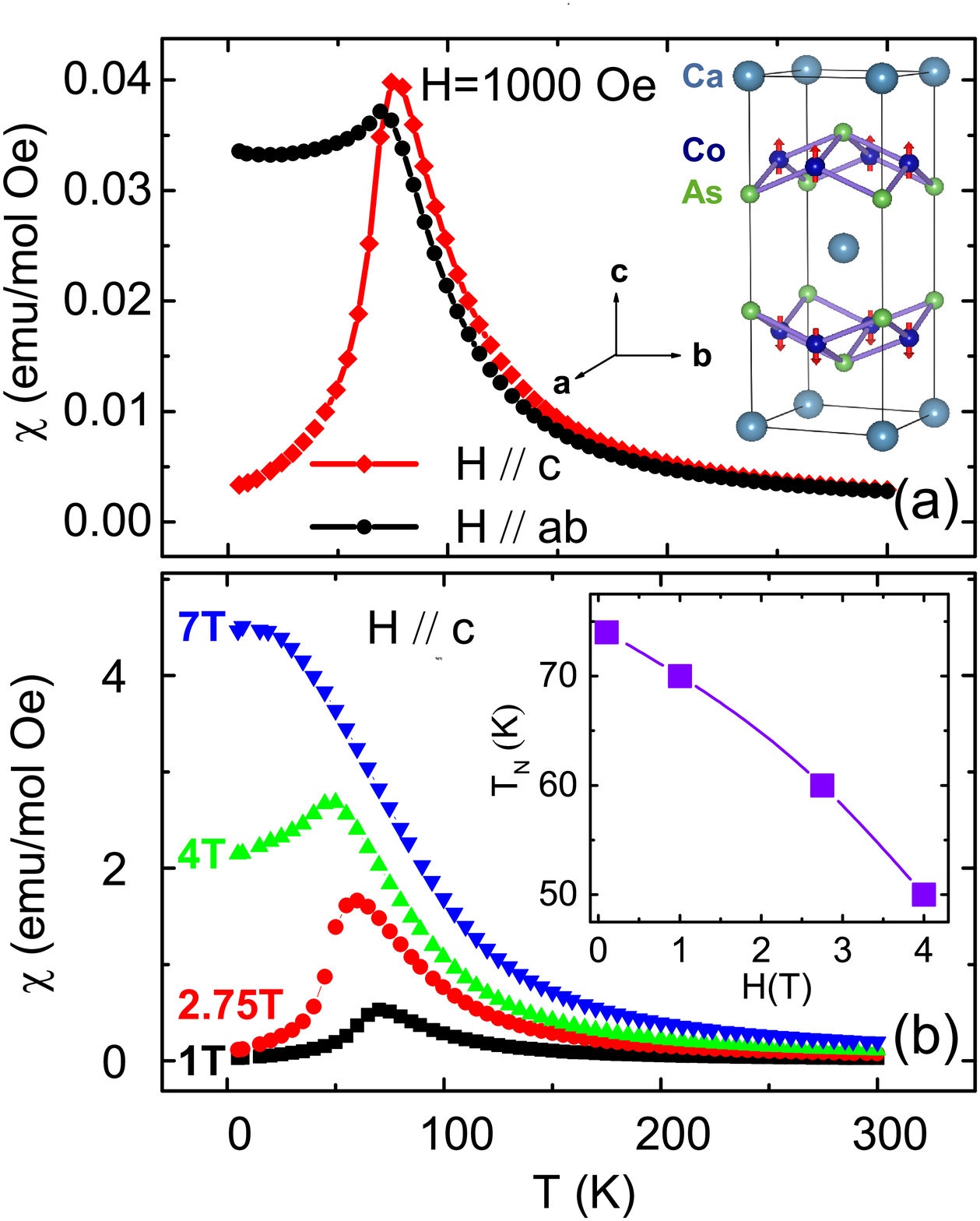}
\caption{(color online) (a) The \emph{M(T)} curves with \emph{H} $\parallel$ \emph{c} (black circle) and \emph{H} $\parallel$ \emph{ab} (red diamond) at \emph{H} = 0.1 T. Inset: The unit cell of CaCo$_{2}$As$_{2}$ single crystal. (b) The \emph{M(T)} curves in different magnetic fields with \emph{H} $\parallel$  \emph{c}. Inset: Magnetic fields dependence of \emph{T$_{N}$}.  }
\label{Fig2}
\end{figure}
%

%
\begin{figure}[tb]
\includegraphics[width=0.9\columnwidth]{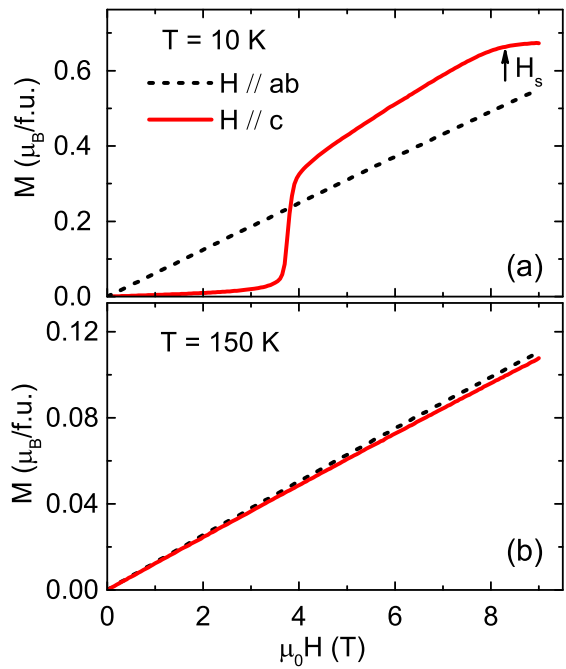}
\caption{(color online) \emph{M(H)} curves for \emph{H} $\parallel$ c and \emph{H} $\parallel$ ab at (a) \emph{T}= 10 K, (b) \emph{T} = 150 K. }
\label{Fig3}
\end{figure}

Fig. 2(a) shows the field cooling (FC) magnetization in the temperature range from 5 to 300 K under magnetic field \emph{H} = 0.1 T in \emph{H} $\parallel$ \emph{c} (diamonds) and \emph{H} $\parallel$ \emph{ab} (circles). For \emph{H} $\parallel$ \emph{c}, the magnetization exhibits a sharp peak at 74 K and drops rapidly upon further cooling. For \emph{H} $\parallel$ \emph{ab}, the magnetization exhibits a plateau below 74 K. We also give a Curie-Weiss fits ($\chi=\frac{C}{T-\theta}$, figure not shown) to the high temperature parts of susceptibility (150 K $\leq T \leq$ 300K ), it gives a Weiss temperature $\theta_{c} = +82.5(3)$ K for $H \parallel c$ and $\theta_{ab} = +71.3(2)$ K for $H \parallel ab$. In general, a positive Weiss temperature means a ferromagnetic coupling between moments. Therefore, the above results may suggest that the Co spins aligned antiferromagnetically along the \emph{c} axis and ferromagnetically in the \emph{ab} plane. The A-type AFM structure were also inferred and confirmed in previous reports.\cite{Cheng2012,Ying2012,Quirinale2013,Anand2014}. The inset in Fig. 2(a) illustrates the unit cell of CaCo$_{2}$As$_{2}$ and the observed magnetic order of Co spins.\cite{Quirinale2013,Anand2014} Both \emph{M(T)} and \emph{$\rho$(T)} measurements confirm the AFM transition at 74 K which is close to 76 K reported by Cheng et al.,\cite{Cheng2012} but higher than the reported value of 52 K by Quirinale et al..\cite{Quirinale2013} We notice that there is controversy between the $T_N$ values by Cheng et al.\cite{Cheng2012} and by Quirinale et al..\cite{Quirinale2013} It has been argued that the lower $T_N$ in Sn-flux samples is a result of Co vacancies. However, considering the fact that in our samples there are also Co vacancies, and the $T_N$ is as high as that in samples prepared by CoAs self-flux. Therefore, we think Co vacancies is not the sole reason for the lower $T_N$,  it may also be related to the actual conditions in the crystal growth, more experiments are needed to clarify it. Fig. 2(b) shows the FC magnetization in different applied fields. It is obvious that \emph{T$_{N}$} peak shifts to lower temperature with increasing magnetic field, and it vanishes at 7 T due to suppression of AFM with applied magnetic field, as shown in the inset of Fig. 2(b).

Figs. 3(a-b) show the \emph{M(H)} curves at \emph{T} = 10 and 150 K for \emph{H} $\parallel$ \emph{c} (solid line) and \emph{H} $\parallel$ \emph{ab} (dashed line). At \emph{T} = 10 K (\emph{H} $\parallel$ \emph{c}), the curve shows an AFM feature at low fields, but \emph{M} suddenly increases around 3.7 T with a sharp step and finally tends to saturate at above 7.5 T. We attributed the sharp increase in \emph{M} to SF transition. We notice that the system shows only one SF transition at 3.7 T. For \emph{H} $\parallel$ \emph{ab}, \emph{M$_{ab}$} increases linearly and no saturation magnetization has been observed with \emph{H} up to 9 T. Meanwhile, at 150 K, which is well above \emph{T$_{N}$}, the \emph{M(H)} curves are essentially linear in \emph{H} $\parallel$ \emph{c} and \emph{H} $\parallel$ \emph{ab} and don't saturate even up to 9 T as shown in Fig. 3(b).

\subsection{Magnetic torque} 

In general, the magnetic torque of a paramagnet with an uniaxial anisotropy in SI units takes the following form,\cite{Xiao2006}
\begin{equation}
\label{PM}
\tau_{PM}=\frac{1}{2}(\chi_{\bot}-\chi_{\|})\mu_{0} H^{2}\sin 2\theta.
\end{equation}
For an uniaxial anistropic antiferromagnet, the free energy \emph{F} of the system can be expressed with the following equation,\cite{Nagamiya1951a,Yosida1951,Nagamiya1955}
\begin{equation}
\label{free energy F}
F=-\frac{1}{2}(\chi_{\bot}\sin^{2}(\phi)+\chi_{\|}\cos^{2}(\phi))\mu_{0}H^{2}+K_{\mu} \sin^{2} (\phi-\theta).
\end{equation}
Since torque \emph{$\tau$} is the angular derivative of free energy \emph{F}, we get\cite{Yosida1951,Nagamiya1955,Weyeneth2011,Watson2014}
\begin{equation}
\label{TORQUE}
\tau_{AFM}=-\frac{\partial F}{\partial \theta}
    =\frac{1}{2}(\chi_{\bot}-\chi_{\|})\mu_{0} H^{2}\frac{\sin 2\theta}{\sqrt{\lambda^{2}-2\lambda \cos 2\theta+1}}.
\end{equation}
Here, we have
\begin{equation}
\label{constant}
\lambda=\left(\frac{H}{H_{SF}}\right)^{2},
\end{equation}
and the SF field
\begin{equation}
\label{Hsf}
H_{SF}=\sqrt{\frac{2K_{\mu}}{\mu_{0}(\chi_{\bot}-\chi_{\|})}}.
\end{equation}
There are two terms of the free energy in Eq.(2): the first term is the magnetic energy and the second is the anisotropy energy, $\chi_{\bot}$ and $\chi_{\|}$ are the spin susceptibilities perpendicular and parallel to the magnetization easy axis, $\phi$ is the angle between the applied magnetic field and spin axis, $\theta$ is the angle between the applied magnetic field and the easy axis, $\mu_{0}$ is the vacuum magnetic permeability, and \emph{K$_{\mu}$} is the anisotropy energy.

%
\begin{figure}[tb]
\includegraphics[width=0.9\columnwidth]{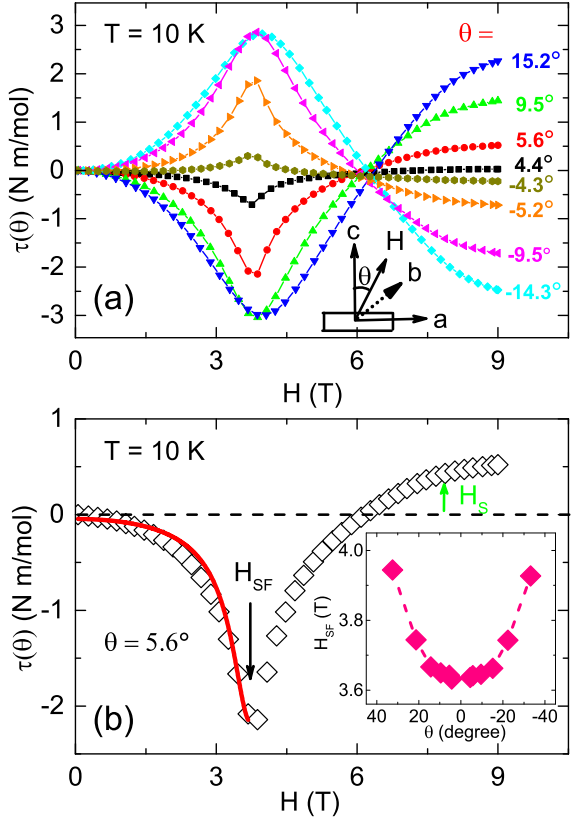}
\caption{(color online) (a) $\tau(H)$ curves of CaCo$_{2}$As$_{2}$ for $\theta$ from $-$14.3$^{\circ}$ to 15.2$^{\circ}$ at \emph{T} = 10 K. (b) $\tau(H)$ curve at $\theta$ = 5.6$^{\circ}$ and the fitting results below $H_{SF}$ by Eq.(3) (red solid line). Inset: Angular-dependent SF transition field \emph{H$_{SF}$} at \emph{T} = 10 K.}
\label{Fig4}
\end{figure}
%

%
\begin{figure}[tb]
\includegraphics[width=0.9\columnwidth]{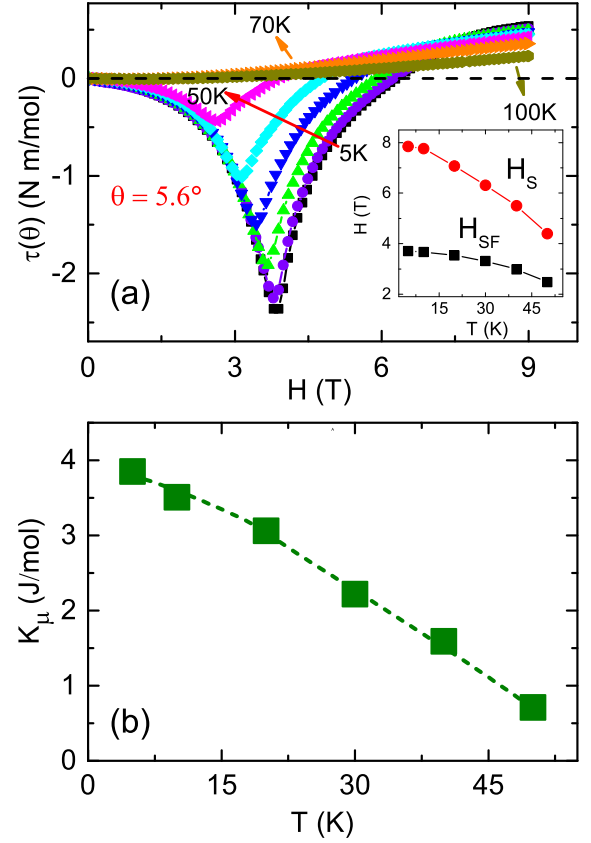}
\caption{(color online) (a) Magnetic field dependence of magnetic torque curves at $\theta$ = 5.6$^{\circ}$ at different temperatures. Inset: Temperature-dependence of $H_{SF}$ and $H_{FM}$ at \emph{T} = 5, 10, 20, 30, 40 and 50 K. (b) \emph{T}-dependent anisotropic energy $K_{\mu}$ obtained from the fitting result by Eq. (3). }
\label{Fig5}
\end{figure}
%

\begin{figure*}[tb]
\includegraphics[width=1.8\columnwidth]{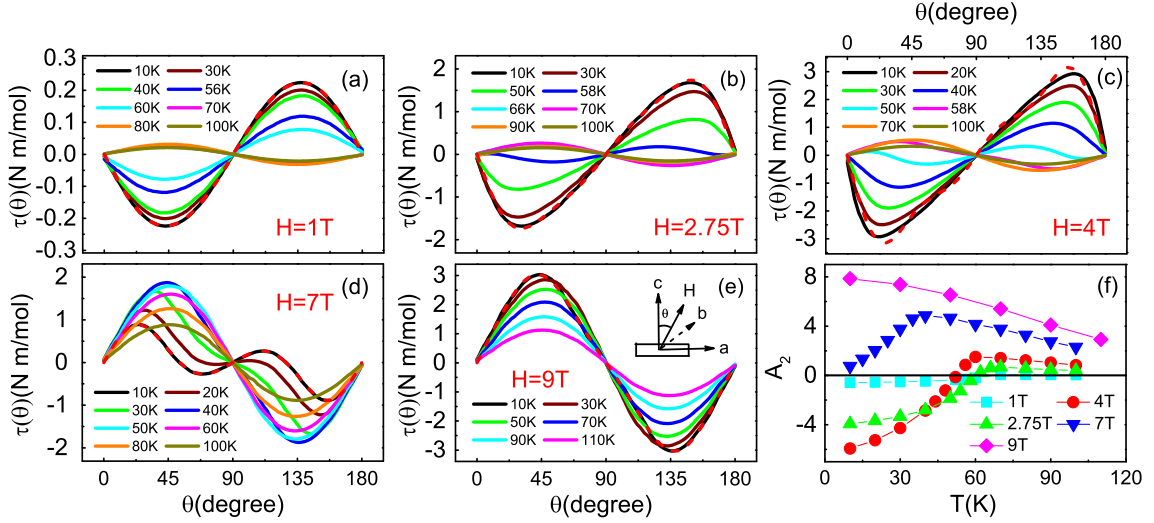}
\caption{ (color online) Angular dependence of magnetic torque of CaCo$_{2}$As$_{2}$ single crystal in different magnetic fields. (a) \emph{H} = 1 T. (b) \emph{H} = 2.75 T. (c) \emph{H} = 4 T. (d) \emph{H} = 7 T. (e) \emph{H} = 9 T. (f) Temperature dependence of the coefficient A$_{2}$ in different magnetic fields.  The red dashed lines at \emph{T} = 10 K in each magnetic field are the fitting results with Eq. (6).}
\label{Fig6}
\end{figure*}

Fig. 4(a) shows the magnetic torque curves at different applied field direction $\theta$ at \emph{T} = 10 K. $\theta$ is the angle between magnetic field \emph{H} and \emph{c}-axis. It shows a sharp dip for $\theta > 0^\circ$  which changes to a peak for $\theta < 0^\circ$ at $H_{SF}$. When \emph{H} $>$ $H_{SF}$, the system changes to a SF phases and the torque amplitude decreases and shows a sign change with increasing \emph{H}. With further increasing \emph{H}, $\tau(H)$ curves deviate from their trend in higher field, and tend to saturate. We attributed this tendency of saturation at \emph{H$_{S}$} to a separation from SF phases to PM state. The SF dip becomes broader as $\theta$ gets away from zero. Moreover, as shown in the inset of Fig. 4(b), $H_{SF}$ slightly increases with increasing tilting angle $\theta$ and shows a minimum when the magnetic field is oriented along the \emph{c} axis (\emph{H} $\parallel$ \emph{c}), this feature indicates that the \emph{c}-axis is the easy axis for the ordered Co spins in this system.\cite{Xu2010}

In order to understand the field-dependent torque, we choose one sharp curve at $\theta = 5.6^{\circ}$ to analyze in detail. As shown in the Fig. 4(b), a dip position (black arrow) and a saturation tendency (green arrow) are observed at \emph{H$_{SF}$} = 3.7 T and \emph{H$_{S}$} = 7.7 T, respectively. Due to Eq. (3) is only right at which the parallel and perpendicular magnetization are proportional to the applied fields, we fit the experimental $\tau(H)$ curves just below $H_{SF}$ and find that the fitting is almost in agreement with the experimental data. With further increasing field, the observed saturation of the torque signal above $H_{S}$ is a result of Co spins aligned along the direction of applied fields in high fields. This tendency of saturation is consistent with the results of \emph{M(H)} as shown in Fig. 3(a).

Next, we will discuss the field dependence of magnetic torque signal for $\theta\sim 5.6^{\circ}$ at different temperatures. As shown in Fig. 5(a), both the amplitude of SF dip and \emph{H$_{SF}$} decrease with increasing temperature. At higher temperatures, the magnetic spins fluctuate around their anisotropy axes and smaller field is required for SF transition, and finally SF dip vanishes above 70 K where a magnetic transition from AFM to PM state occurs. The AFM to PM transition temperature $T_N$ deduced from \emph{$\tau$(T)} curves is consistent with that as determined from \emph{M(T)}. $H_{S}$ also shows a decreasing trend with increasing temperature as shown in the inset of Fig. 5(a). We have also fitted these \emph{$\tau$(H)} curves below $H_{SF}$ by using Eq. (3), the fitted parameters $K_{\mu}$ for different temperatures are plotted in Fig. 5(b). It shows a decreasing trend with increasing temperature due to decrease in AFM anisotropy energy with temperature.

Figs. 6(a-e) show the angular dependent torque curves in the temperature region of 10-100 K under different applied fields. It is evident that all the $\tau(\theta)$ curves with \emph{H} = 1 T exhibits a sinusoidal wave shape and an extremum (maxima or minima) around $\theta = 45^{\circ}$. The $\tau(\theta)$ curves show a symmetrical behavior at $90^{\circ}$, therefore we will discuss only the first part that lies in range $0^{\circ}$ $\leq$ $\theta$ $\leq$ $90^{\circ}$. The amplitude of the minima decreases with increasing temperature and becomes positive (a sign change) around 70 K, which is consistent with the $T_N$ = 70 K at \emph{H} = 1 T as determined from \emph{M(T)} (Fig. 2(b)). Therefore, a sign change in the $\tau(\theta)$ curves with increasing temperature is due to the magnetic transition from AFM to PM state. Similar changes were also observed for \emph{H} = 2.75 T and \emph{H} = 4 T as shown in Fig. 6(b) and 6(c). However, the minima in this two fields shifts away from $\theta = 45^{\circ}$ and the sinusoidal wave shape transfers into the saw-tooth waveform, which are attributed to the SF transition between 2.75 and 4 T. At \emph{H} = 7 T (Fig. 6(d)), the lower temperature curves exhibit a very different shape as compared to curves at higher temperature, which can arise from a transition from SF phases to PM state. On further increasing magnetic field to 9 T (Fig. 6(e)), the low temperature $\tau(\theta)$ curve at 10 K exhibits only a positive peak in $0^{\circ}$ $\leq$ $\theta$ $\leq$ $90^{\circ}$, because the system is in the field induced saturated PM state at 9 T.  The amplitude of the peak in $\tau(\theta)$ curves decreases with increasing temperature but exhibits only positive $\tau(\theta)$ in the whole temperature region ( 10 K $\leq T \leq$ 100 K). The decrease of peak amplitude with temperature is due to increasing thermal fluctuations at high temperature which disturbs the Co spins of saturated PM . The magnetic transitions from AFM to PM as determined from $\tau(\theta)$ analysis are in agreement with the \emph{M(T)} results.

The angular dependent magnetic torque can be expanded as\cite{Uozaki2000}
\begin{equation}
\label{expansion}
\tau= A_{2}\sin 2\theta+A_{4}\sin 4\theta+A_{6}\sin 6\theta+\ldots,
\end{equation}
where the first term is the first order contribution as described in Eq. (1), and the second and third terms are higher order corrections. The magnetic torque measured as a function of angle at different fields are fitted by Eq. (6) (the red dashed line at \emph{T} = 10 K in Figs. 6(a)-6(e)). The typical fitting results of \emph{A$_{2}$} is shown in Fig. 6(f). For applied fields between 1 T and 4 T, as the temperature increases, \emph{A$_{2}$} shows a sign change, because of the magnetic transition from AFM to PM. The transition temperature decreases with increasing field and shifts from 70 K to 54 K as the field increases from 1 T to 4 T, which is in agreement with the dc magnetization results (Fig. 2(b)).

%
\begin{figure}[tb]
\includegraphics[width=0.8\columnwidth]{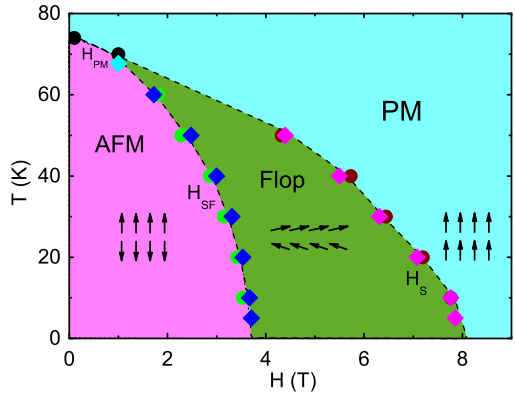}
\caption{ (color online) \emph{T-H} phase diagram concluded from magnetization (circles) at $H \parallel c$ and magnetic torque (diamonds) measurements at $\theta = 5.6^{\circ}$ for CaCo$_{2}$As$_{2}$. The curves are guides to the eye. The collinear AFM structure, SF phases and the arrangement of Co moments above $H_S$ with $H \parallel c$ are shown with the corresponding arrows in the figure.}
\label{Fig7}
\end{figure}

Fig. 7 summarizes the detailed magnetic phase diagram of CaCo$_{2}$As$_{2}$ single crystal as derived from magnetization (circles) and magnetic torque (diamonds) measurements. We define following terms as: \emph{H$_{SF}$ }is the SF onset field corresponds to the rapid increase in \emph{M(H)} at $H \parallel c$ and the dip position in $\tau(H)$ curves at $\theta$ = 5.6$^{\circ}$, the AFM transition temperature is extracted from $\chi(T)$ and $\tau(T)$ data. As shown in the phase diagram, below $T_N$, the magnetic structure is a collinear A-type AFM at low field. With increasing magnetic field, the SF transition occurs at \emph{H$_{SF}$} and then all Co spins tend to align along the direction of the external field above \emph{H$_{S}$}. The arrows in the phase diagram represent the corresponding collinear AFM structure, SF phases and the arrangement of Co moments above $H_S$, respectively. Meanwhile, above $H_{S}$, as the temperature increases, the Co moments will be disturbed by the increasing thermal fluctuation at high temperature and the system may have a gradually variation to unsaturated PM state. Through a combination of magnetization and magnetic torque measurements, we provide a detailed and systematical way to analyze the metamagnetic transition especially the spin-flop transition.

\section{CONCLUSIONS}

In conclusion, we have systematically investigated the magnetic properties of CaCo$_{2}$As$_{2}$ single crystal through dc magnetization and torque magnetometry. An A-type AFM ground state with $T_N$ = 74 K and with easy axis being aligned along the c axis are inferred from magnetization and magnetic torque measurements. For $H \parallel c$, \emph{M(H)} and $\tau(H)$ data reveals a spin-flop transition below $T_N$. Together with magnetization measurements, a more detailed analysis about the spin-flop transition and magnetic phase diagram is given by torque measurements.

%
%

\begin{acknowledgments}
 This research was supported by the Natural Science Foundation of China (Grant Nos. 11104335, 91121004, 11274237, 51228201, and 11004238) and the Ministry of Science and Technology of China (973 Projects Nos. 2009CB929102, 2011CBA00107, 2012CB821400, 2012CB921302, and 2015CB921303). The authors also acknowledge the support of the Priority Academic Program Development of Jiangsu Higher Education Institutions (PAPD).
\end{acknowledgments}
%


\begin{thebibliography}{41}%
\makeatletter
\providecommand \@ifxundefined [1]{%
 \@ifx{#1\undefined}
}%
\providecommand \@ifnum [1]{%
 \ifnum #1\expandafter \@firstoftwo
 \else \expandafter \@secondoftwo
 \fi
}%
\providecommand \@ifx [1]{%
 \ifx #1\expandafter \@firstoftwo
 \else \expandafter \@secondoftwo
 \fi
}%
\providecommand \natexlab [1]{#1}%
\providecommand \enquote  [1]{``#1''}%
\providecommand \bibnamefont  [1]{#1}%
\providecommand \bibfnamefont [1]{#1}%
\providecommand \citenamefont [1]{#1}%
\providecommand \href@noop [0]{\@secondoftwo}%
\providecommand \href [0]{\begingroup \@sanitize@url \@href}%
\providecommand \@href[1]{\@@startlink{#1}\@@href}%
\providecommand \@@href[1]{\endgroup#1\@@endlink}%
\providecommand \@sanitize@url [0]{\catcode `\\12\catcode `\$12\catcode
  `\&12\catcode `\#12\catcode `\^12\catcode `\_12\catcode `\%12\relax}%
\providecommand \@@startlink[1]{}%
\providecommand \@@endlink[0]{}%
\providecommand \url  [0]{\begingroup\@sanitize@url \@url }%
\providecommand \@url [1]{\endgroup\@href {#1}{\urlprefix }}%
\providecommand \urlprefix  [0]{URL }%
\providecommand \Eprint [0]{\href }%
\providecommand \doibase [0]{http://dx.doi.org/}%
\providecommand \selectlanguage [0]{\@gobble}%
\providecommand \bibinfo  [0]{\@secondoftwo}%
\providecommand \bibfield  [0]{\@secondoftwo}%
\providecommand \translation [1]{[#1]}%
\providecommand \BibitemOpen [0]{}%
\providecommand \bibitemStop [0]{}%
\providecommand \bibitemNoStop [0]{.\EOS\space}%
\providecommand \EOS [0]{\spacefactor3000\relax}%
\providecommand \BibitemShut  [1]{\csname bibitem#1\endcsname}%
\let\auto@bib@innerbib\@empty
\bibitem [{\citenamefont {N\'{e}el}(1952)}]{Neel1952}%
  \BibitemOpen
  \bibfield  {author} {\bibinfo {author} {\bibfnamefont {L.}~\bibnamefont
  {N\'{e}el}},\ }\href {http://stacks.iop.org/0370-1298/65/i=11/a=301} {\bibfield
  {journal} {\bibinfo  {journal} {Proc. Phys. Soc. A}\ }\textbf {\bibinfo
  {volume} {65}},\ \bibinfo {pages} {869} (\bibinfo {year} {1952})}\BibitemShut
  {NoStop}%
\bibitem [{\citenamefont {Fries}\ \emph {et~al.}(1997)\citenamefont {Fries},
  \citenamefont {Shapira}, \citenamefont {Palacio}, \citenamefont {Mor\'on},
  \citenamefont {McIntyre}, \citenamefont {Kershaw}, \citenamefont {Wold},\
  and\ \citenamefont {McNiff}}]{Fries1997}%
  \BibitemOpen
  \bibfield  {author} {\bibinfo {author} {\bibfnamefont {T.}~\bibnamefont
  {Fries}}, \bibinfo {author} {\bibfnamefont {Y.}~\bibnamefont {Shapira}},
  \bibinfo {author} {\bibfnamefont {F.}~\bibnamefont {Palacio}}, \bibinfo
  {author} {\bibfnamefont {M.~C.}\ \bibnamefont {Mor\'on}}, \bibinfo {author}
  {\bibfnamefont {G.~J.}\ \bibnamefont {McIntyre}}, \bibinfo {author}
  {\bibfnamefont {R.}~\bibnamefont {Kershaw}}, \bibinfo {author} {\bibfnamefont
  {A.}~\bibnamefont {Wold}}, \ and\ \bibinfo {author} {\bibfnamefont {E.~J.}\
  \bibnamefont {McNiff}},\ }\href {\doibase 10.1103/PhysRevB.56.5424}
  {\bibfield  {journal} {\bibinfo  {journal} {Phys. Rev. B}\ }\textbf {\bibinfo
  {volume} {56}},\ \bibinfo {pages} {5424} (\bibinfo {year}
  {1997})}\BibitemShut {NoStop}%
\bibitem [{\citenamefont {Uozaki}\ \emph {et~al.}(2000)\citenamefont {Uozaki},
  \citenamefont {Sasaki}, \citenamefont {Endo},\ and\ \citenamefont
  {Toyota}}]{Uozaki2000}%
  \BibitemOpen
  \bibfield  {author} {\bibinfo {author} {\bibfnamefont {H.}~\bibnamefont
  {Uozaki}}, \bibinfo {author} {\bibfnamefont {T.}~\bibnamefont {Sasaki}},
  \bibinfo {author} {\bibfnamefont {S.}~\bibnamefont {Endo}}, \ and\ \bibinfo
  {author} {\bibfnamefont {N.}~\bibnamefont {Toyota}},\ }\href {\doibase
  10.1143/JPSJ.69.2759} {\bibfield  {journal} {\bibinfo  {journal} {J. Phys.
  Soc. Jpn.}\ }\textbf {\bibinfo {volume} {69}},\ \bibinfo {pages} {2759}
  (\bibinfo {year} {2000})}\BibitemShut {NoStop}%
\bibitem [{\citenamefont {Sasaki}\ \emph {et~al.}(2001)\citenamefont {Sasaki},
  \citenamefont {Uozaki}, \citenamefont {Endo},\ and\ \citenamefont
  {Toyota}}]{Sasaki2001}%
  \BibitemOpen
  \bibfield  {author} {\bibinfo {author} {\bibfnamefont {T.}~\bibnamefont
  {Sasaki}}, \bibinfo {author} {\bibfnamefont {H.}~\bibnamefont {Uozaki}},
  \bibinfo {author} {\bibfnamefont {S.}~\bibnamefont {Endo}}, \ and\ \bibinfo
  {author} {\bibfnamefont {N.}~\bibnamefont {Toyota}},\ }\href {\doibase
  http://dx.doi.org/10.1016/S0379-6779(00)00775-X} {\bibfield  {journal}
  {\bibinfo  {journal} {Synth. Metals}\ }\textbf {\bibinfo {volume} {120}},\
  \bibinfo {pages} {759 } (\bibinfo {year} {2001})}\BibitemShut {NoStop}%
\bibitem [{\citenamefont {Wernsdorfer}\ \emph {et~al.}(2002)\citenamefont
  {Wernsdorfer}, \citenamefont {Aliaga-Alcalde}, \citenamefont {Hendrickson},\
  and\ \citenamefont {Christou}}]{Wernsdorfer2002}%
  \BibitemOpen
  \bibfield  {author} {\bibinfo {author} {\bibfnamefont {W.}~\bibnamefont
  {Wernsdorfer}}, \bibinfo {author} {\bibfnamefont {N.}~\bibnamefont
  {Aliaga-Alcalde}}, \bibinfo {author} {\bibfnamefont {D.~N.}\ \bibnamefont
  {Hendrickson}}, \ and\ \bibinfo {author} {\bibfnamefont {G.}~\bibnamefont
  {Christou}},\ }\href {http://dx.doi.org/10.1038/416406a} {\bibfield
  {journal} {\bibinfo  {journal} {Nature}\ }\textbf {\bibinfo {volume} {416}},\
  \bibinfo {pages} {406} (\bibinfo {year} {2002})}\BibitemShut {NoStop}%
\bibitem [{\citenamefont {Kawamoto}\ \emph {et~al.}(2008)\citenamefont
  {Kawamoto}, \citenamefont {Bando}, \citenamefont {Mori}, \citenamefont
  {Konoike}, \citenamefont {Takahide}, \citenamefont {Terashima}, \citenamefont
  {Uji}, \citenamefont {Takimiya},\ and\ \citenamefont
  {Otsubo}}]{Kawamoto2008}%
  \BibitemOpen
  \bibfield  {author} {\bibinfo {author} {\bibfnamefont {T.}~\bibnamefont
  {Kawamoto}}, \bibinfo {author} {\bibfnamefont {Y.}~\bibnamefont {Bando}},
  \bibinfo {author} {\bibfnamefont {T.}~\bibnamefont {Mori}}, \bibinfo {author}
  {\bibfnamefont {T.}~\bibnamefont {Konoike}}, \bibinfo {author} {\bibfnamefont
  {Y.}~\bibnamefont {Takahide}}, \bibinfo {author} {\bibfnamefont
  {T.}~\bibnamefont {Terashima}}, \bibinfo {author} {\bibfnamefont
  {S.}~\bibnamefont {Uji}}, \bibinfo {author} {\bibfnamefont {K.}~\bibnamefont
  {Takimiya}}, \ and\ \bibinfo {author} {\bibfnamefont {T.}~\bibnamefont
  {Otsubo}},\ }\href {\doibase 10.1103/PhysRevB.77.224506} {\bibfield
  {journal} {\bibinfo  {journal} {Phys. Rev. B}\ }\textbf {\bibinfo {volume}
  {77}},\ \bibinfo {pages} {224506} (\bibinfo {year} {2008})}\BibitemShut
  {NoStop}%
\bibitem [{\citenamefont {Fitzsimmons}\ \emph {et~al.}(2000)\citenamefont
  {Fitzsimmons}, \citenamefont {Yashar}, \citenamefont {Leighton},
  \citenamefont {Schuller}, \citenamefont {Nogu\'es}, \citenamefont
  {Majkrzak},\ and\ \citenamefont {Dura}}]{Fitzsimmons2000}%
  \BibitemOpen
  \bibfield  {author} {\bibinfo {author} {\bibfnamefont {M.~R.}\ \bibnamefont
  {Fitzsimmons}}, \bibinfo {author} {\bibfnamefont {P.}~\bibnamefont {Yashar}},
  \bibinfo {author} {\bibfnamefont {C.}~\bibnamefont {Leighton}}, \bibinfo
  {author} {\bibfnamefont {I.~K.}\ \bibnamefont {Schuller}}, \bibinfo {author}
  {\bibfnamefont {J.}~\bibnamefont {Nogu\'es}}, \bibinfo {author}
  {\bibfnamefont {C.~F.}\ \bibnamefont {Majkrzak}}, \ and\ \bibinfo {author}
  {\bibfnamefont {J.~A.}\ \bibnamefont {Dura}},\ }\href {\doibase
  10.1103/PhysRevLett.84.3986} {\bibfield  {journal} {\bibinfo  {journal}
  {Phys. Rev. Lett.}\ }\textbf {\bibinfo {volume} {84}},\ \bibinfo {pages}
  {3986} (\bibinfo {year} {2000})}\BibitemShut {NoStop}%
\bibitem [{\citenamefont {Reehuis}\ and\ \citenamefont
  {Jeitschko}(1990)}]{Reehuis1990}%
  \BibitemOpen
  \bibfield  {author} {\bibinfo {author} {\bibfnamefont {M.}~\bibnamefont
  {Reehuis}}\ and\ \bibinfo {author} {\bibfnamefont {W.}~\bibnamefont
  {Jeitschko}},\ }\href {\doibase
  http://dx.doi.org/10.1016/0022-3697(90)90039-I} {\bibfield  {journal}
  {\bibinfo  {journal} {J. Phys. Chem. Solids}\ }\textbf {\bibinfo {volume}
  {51}},\ \bibinfo {pages} {961 } (\bibinfo {year} {1990})}\BibitemShut
  {NoStop}%
\bibitem [{\citenamefont {Reehuis}\ \emph {et~al.}(1993)\citenamefont
  {Reehuis}, \citenamefont {Brown}, \citenamefont {Jeitschko}, \citenamefont
  {Möller},\ and\ \citenamefont {Vomhof}}]{Reehuis1993}%
  \BibitemOpen
  \bibfield  {author} {\bibinfo {author} {\bibfnamefont {M.}~\bibnamefont
  {Reehuis}}, \bibinfo {author} {\bibfnamefont {P.}~\bibnamefont {Brown}},
  \bibinfo {author} {\bibfnamefont {W.}~\bibnamefont {Jeitschko}}, \bibinfo
  {author} {\bibfnamefont {M.}~\bibnamefont {Möller}}, \ and\ \bibinfo
  {author} {\bibfnamefont {T.}~\bibnamefont {Vomhof}},\ }\href {\doibase
  http://dx.doi.org/10.1016/0022-3697(93)90330-T} {\bibfield  {journal}
  {\bibinfo  {journal} {J. Phys. Chem. Solids}\ }\textbf {\bibinfo {volume}
  {54}},\ \bibinfo {pages} {469 } (\bibinfo {year} {1993})}\BibitemShut
  {NoStop}%
\bibitem [{\citenamefont {Reehuis}\ \emph {et~al.}(1994)\citenamefont
  {Reehuis}, \citenamefont {Ritter}, \citenamefont {Ballou},\ and\
  \citenamefont {Jeitschko}}]{Reehuis1994}%
  \BibitemOpen
  \bibfield  {author} {\bibinfo {author} {\bibfnamefont {M.}~\bibnamefont
  {Reehuis}}, \bibinfo {author} {\bibfnamefont {C.}~\bibnamefont {Ritter}},
  \bibinfo {author} {\bibfnamefont {R.}~\bibnamefont {Ballou}}, \ and\ \bibinfo
  {author} {\bibfnamefont {W.}~\bibnamefont {Jeitschko}},\ }\href {\doibase
  http://dx.doi.org/10.1016/0304-8853(94)90402-2} {\bibfield  {journal}
  {\bibinfo  {journal} {J. Magn. Magn. Mater.}\ }\textbf {\bibinfo {volume}
  {138}},\ \bibinfo {pages} {85 } (\bibinfo {year} {1994})}\BibitemShut
  {NoStop}%
\bibitem [{\citenamefont {Reehuis}\ \emph {et~al.}(1998)\citenamefont
  {Reehuis}, \citenamefont {Jeitschko}, \citenamefont {Kotzyba}, \citenamefont
  {Zimmer},\ and\ \citenamefont {Hu}}]{Reehuis1998}%
  \BibitemOpen
  \bibfield  {author} {\bibinfo {author} {\bibfnamefont {M.}~\bibnamefont
  {Reehuis}}, \bibinfo {author} {\bibfnamefont {W.}~\bibnamefont {Jeitschko}},
  \bibinfo {author} {\bibfnamefont {G.}~\bibnamefont {Kotzyba}}, \bibinfo
  {author} {\bibfnamefont {B.}~\bibnamefont {Zimmer}}, \ and\ \bibinfo {author}
  {\bibfnamefont {X.}~\bibnamefont {Hu}},\ }\href {\doibase
  http://dx.doi.org/10.1016/S0925-8388(97)00486-6} {\bibfield  {journal}
  {\bibinfo  {journal} {J. Alloys Comp.}\ }\textbf {\bibinfo {volume} {266}},\
  \bibinfo {pages} {54 } (\bibinfo {year} {1998})}\BibitemShut {NoStop}%
\bibitem [{\citenamefont {Rotter}\ \emph
  {et~al.}(2008{\natexlab{a}})\citenamefont {Rotter}, \citenamefont {Tegel},
  \citenamefont {Johrendt}, \citenamefont {Schellenberg}, \citenamefont
  {Hermes},\ and\ \citenamefont {P\"ottgen}}]{Rotter2008}%
  \BibitemOpen
  \bibfield  {author} {\bibinfo {author} {\bibfnamefont {M.}~\bibnamefont
  {Rotter}}, \bibinfo {author} {\bibfnamefont {M.}~\bibnamefont {Tegel}},
  \bibinfo {author} {\bibfnamefont {D.}~\bibnamefont {Johrendt}}, \bibinfo
  {author} {\bibfnamefont {I.}~\bibnamefont {Schellenberg}}, \bibinfo {author}
  {\bibfnamefont {W.}~\bibnamefont {Hermes}}, \ and\ \bibinfo {author}
  {\bibfnamefont {R.}~\bibnamefont {P\"ottgen}},\ }\href {\doibase
  10.1103/PhysRevB.78.020503} {\bibfield  {journal} {\bibinfo  {journal} {Phys.
  Rev. B}\ }\textbf {\bibinfo {volume} {78}},\ \bibinfo {pages} {020503}
  (\bibinfo {year} {2008}{\natexlab{a}})}\BibitemShut {NoStop}%
\bibitem [{\citenamefont {Rotter}\ \emph
  {et~al.}(2008{\natexlab{b}})\citenamefont {Rotter}, \citenamefont {Tegel},\
  and\ \citenamefont {Johrendt}}]{Rotter2008a}%
  \BibitemOpen
  \bibfield  {author} {\bibinfo {author} {\bibfnamefont {M.}~\bibnamefont
  {Rotter}}, \bibinfo {author} {\bibfnamefont {M.}~\bibnamefont {Tegel}}, \
  and\ \bibinfo {author} {\bibfnamefont {D.}~\bibnamefont {Johrendt}},\ }\href
  {\doibase 10.1103/PhysRevLett.101.107006} {\bibfield  {journal} {\bibinfo
  {journal} {Phys. Rev. Lett.}\ }\textbf {\bibinfo {volume} {101}},\ \bibinfo
  {pages} {107006} (\bibinfo {year} {2008}{\natexlab{b}})}\BibitemShut
  {NoStop}%
\bibitem [{\citenamefont {de~la Cruz}\ \emph {et~al.}(2008)\citenamefont {de~la
  Cruz}, \citenamefont {Huang}, \citenamefont {Lynn}, \citenamefont {Li},
  \citenamefont {II}, \citenamefont {Zarestky}, \citenamefont {Mook},
  \citenamefont {Chen}, \citenamefont {Luo}, \citenamefont {Wang},\ and\
  \citenamefont {Dai}}]{Cruz2008}%
  \BibitemOpen
  \bibfield  {author} {\bibinfo {author} {\bibfnamefont {C.}~\bibnamefont
  {de~la Cruz}}, \bibinfo {author} {\bibfnamefont {Q.}~\bibnamefont {Huang}},
  \bibinfo {author} {\bibfnamefont {J.~W.}\ \bibnamefont {Lynn}}, \bibinfo
  {author} {\bibfnamefont {J.}~\bibnamefont {Li}}, \bibinfo {author}
  {\bibfnamefont {W.}\ \bibnamefont {Ratcliff II}}, \bibinfo {author} {\bibfnamefont
  {J.~L.}\ \bibnamefont {Zarestky}}, \bibinfo {author} {\bibfnamefont {H.~A.}\
  \bibnamefont {Mook}}, \bibinfo {author} {\bibfnamefont {G.~F.}\ \bibnamefont
  {Chen}}, \bibinfo {author} {\bibfnamefont {J.~L.}\ \bibnamefont {Luo}},
  \bibinfo {author} {\bibfnamefont {N.~L.}\ \bibnamefont {Wang}}, \ and\
  \bibinfo {author} {\bibfnamefont {P.}~\bibnamefont {Dai}},\ }\href
  {http://dx.doi.org/10.1038/nature07057} {\bibfield  {journal} {\bibinfo
  {journal} {Nature}\ }\textbf {\bibinfo {volume} {453}},\ \bibinfo {pages}
  {899} (\bibinfo {year} {2008})}\BibitemShut {NoStop}%
\bibitem [{\citenamefont {Fernandes}\ \emph {et~al.}(2010)\citenamefont
  {Fernandes}, \citenamefont {Pratt}, \citenamefont {Tian}, \citenamefont
  {Zarestky}, \citenamefont {Kreyssig}, \citenamefont {Nandi}, \citenamefont
  {Kim}, \citenamefont {Thaler}, \citenamefont {Ni}, \citenamefont {Canfield},
  \citenamefont {McQueeney}, \citenamefont {Schmalian},\ and\ \citenamefont
  {Goldman}}]{Fernandes2010}%
  \BibitemOpen
  \bibfield  {author} {\bibinfo {author} {\bibfnamefont {R.~M.}\ \bibnamefont
  {Fernandes}}, \bibinfo {author} {\bibfnamefont {D.~K.}\ \bibnamefont
  {Pratt}}, \bibinfo {author} {\bibfnamefont {W.}~\bibnamefont {Tian}},
  \bibinfo {author} {\bibfnamefont {J.}~\bibnamefont {Zarestky}}, \bibinfo
  {author} {\bibfnamefont {A.}~\bibnamefont {Kreyssig}}, \bibinfo {author}
  {\bibfnamefont {S.}~\bibnamefont {Nandi}}, \bibinfo {author} {\bibfnamefont
  {M.~G.}\ \bibnamefont {Kim}}, \bibinfo {author} {\bibfnamefont
  {A.}~\bibnamefont {Thaler}}, \bibinfo {author} {\bibfnamefont
  {N.}~\bibnamefont {Ni}}, \bibinfo {author} {\bibfnamefont {P.~C.}\
  \bibnamefont {Canfield}}, \bibinfo {author} {\bibfnamefont {R.~J.}\
  \bibnamefont {McQueeney}}, \bibinfo {author} {\bibfnamefont {J.}~\bibnamefont
  {Schmalian}}, \ and\ \bibinfo {author} {\bibfnamefont {A.~I.}\ \bibnamefont
  {Goldman}},\ }\href {\doibase 10.1103/PhysRevB.81.140501} {\bibfield
  {journal} {\bibinfo  {journal} {Phys. Rev. B}\ }\textbf {\bibinfo {volume}
  {81}},\ \bibinfo {pages} {140501} (\bibinfo {year} {2010})}\BibitemShut
  {NoStop}%
\bibitem [{\citenamefont {Shermadini}\ \emph {et~al.}(2011)\citenamefont
  {Shermadini}, \citenamefont {Krzton-Maziopa}, \citenamefont {Bendele},
  \citenamefont {Khasanov}, \citenamefont {Luetkens}, \citenamefont {Conder},
  \citenamefont {Pomjakushina}, \citenamefont {Weyeneth}, \citenamefont
  {Pomjakushin}, \citenamefont {Bossen},\ and\ \citenamefont
  {Amato}}]{Shermadini2011}%
  \BibitemOpen
  \bibfield  {author} {\bibinfo {author} {\bibfnamefont {Z.}~\bibnamefont
  {Shermadini}}, \bibinfo {author} {\bibfnamefont {A.}~\bibnamefont
  {Krzton-Maziopa}}, \bibinfo {author} {\bibfnamefont {M.}~\bibnamefont
  {Bendele}}, \bibinfo {author} {\bibfnamefont {R.}~\bibnamefont {Khasanov}},
  \bibinfo {author} {\bibfnamefont {H.}~\bibnamefont {Luetkens}}, \bibinfo
  {author} {\bibfnamefont {K.}~\bibnamefont {Conder}}, \bibinfo {author}
  {\bibfnamefont {E.}~\bibnamefont {Pomjakushina}}, \bibinfo {author}
  {\bibfnamefont {S.}~\bibnamefont {Weyeneth}}, \bibinfo {author}
  {\bibfnamefont {V.}~\bibnamefont {Pomjakushin}}, \bibinfo {author}
  {\bibfnamefont {O.}~\bibnamefont {Bossen}}, \ and\ \bibinfo {author}
  {\bibfnamefont {A.}~\bibnamefont {Amato}},\ }\href {\doibase
  10.1103/PhysRevLett.106.117602} {\bibfield  {journal} {\bibinfo  {journal}
  {Phys. Rev. Lett.}\ }\textbf {\bibinfo {volume} {106}},\ \bibinfo {pages}
  {117602} (\bibinfo {year} {2011})}\BibitemShut {NoStop}%
\bibitem [{\citenamefont {Mazin}\ \emph {et~al.}(2008)\citenamefont {Mazin},
  \citenamefont {Singh}, \citenamefont {Johannes},\ and\ \citenamefont
  {Du}}]{Mazin2008}%
  \BibitemOpen
  \bibfield  {author} {\bibinfo {author} {\bibfnamefont {I.~I.}\ \bibnamefont
  {Mazin}}, \bibinfo {author} {\bibfnamefont {D.~J.}\ \bibnamefont {Singh}},
  \bibinfo {author} {\bibfnamefont {M.~D.}\ \bibnamefont {Johannes}}, \ and\
  \bibinfo {author} {\bibfnamefont {M.~H.}\ \bibnamefont {Du}},\ }\href
  {\doibase 10.1103/PhysRevLett.101.057003} {\bibfield  {journal} {\bibinfo
  {journal} {Phys. Rev. Lett.}\ }\textbf {\bibinfo {volume} {101}},\ \bibinfo
  {pages} {057003} (\bibinfo {year} {2008})}\BibitemShut {NoStop}%
\bibitem [{\citenamefont {Dong}\ \emph {et~al.}(2010)\citenamefont {Dong},
  \citenamefont {Zhou}, \citenamefont {Guan}, \citenamefont {Zhang},
  \citenamefont {Dai}, \citenamefont {Qiu}, \citenamefont {Wang}, \citenamefont
  {He}, \citenamefont {Chen},\ and\ \citenamefont {Li}}]{Dong2010}%
  \BibitemOpen
  \bibfield  {author} {\bibinfo {author} {\bibfnamefont {J.~K.}\ \bibnamefont
  {Dong}}, \bibinfo {author} {\bibfnamefont {S.~Y.}\ \bibnamefont {Zhou}},
  \bibinfo {author} {\bibfnamefont {T.~Y.}\ \bibnamefont {Guan}}, \bibinfo
  {author} {\bibfnamefont {H.}~\bibnamefont {Zhang}}, \bibinfo {author}
  {\bibfnamefont {Y.~F.}\ \bibnamefont {Dai}}, \bibinfo {author} {\bibfnamefont
  {X.}~\bibnamefont {Qiu}}, \bibinfo {author} {\bibfnamefont {X.~F.}\
  \bibnamefont {Wang}}, \bibinfo {author} {\bibfnamefont {Y.}~\bibnamefont
  {He}}, \bibinfo {author} {\bibfnamefont {X.~H.}\ \bibnamefont {Chen}}, \ and\
  \bibinfo {author} {\bibfnamefont {S.~Y.}\ \bibnamefont {Li}},\ }\href
  {\doibase 10.1103/PhysRevLett.104.087005} {\bibfield  {journal} {\bibinfo
  {journal} {Phys. Rev. Lett.}\ }\textbf {\bibinfo {volume} {104}},\ \bibinfo
  {pages} {087005} (\bibinfo {year} {2010})}\BibitemShut {NoStop}%
\bibitem [{\citenamefont {Torikachvili}\ \emph {et~al.}(2008)\citenamefont
  {Torikachvili}, \citenamefont {Bud'ko}, \citenamefont {Ni},\ and\
  \citenamefont {Canfield}}]{Torikachvili2008}%
  \BibitemOpen
  \bibfield  {author} {\bibinfo {author} {\bibfnamefont {M.~S.}\ \bibnamefont
  {Torikachvili}}, \bibinfo {author} {\bibfnamefont {S.~L.}\ \bibnamefont
  {Bud'ko}}, \bibinfo {author} {\bibfnamefont {N.}~\bibnamefont {Ni}}, \ and\
  \bibinfo {author} {\bibfnamefont {P.~C.}\ \bibnamefont {Canfield}},\ }\href
  {\doibase 10.1103/PhysRevLett.101.057006} {\bibfield  {journal} {\bibinfo
  {journal} {Phys. Rev. Lett.}\ }\textbf {\bibinfo {volume} {101}},\ \bibinfo
  {pages} {057006} (\bibinfo {year} {2008})}\BibitemShut {NoStop}%
\bibitem [{\citenamefont {Kreyssig}\ \emph {et~al.}(2008)\citenamefont
  {Kreyssig}, \citenamefont {Green}, \citenamefont {Lee}, \citenamefont
  {Samolyuk}, \citenamefont {Zajdel}, \citenamefont {Lynn}, \citenamefont
  {Bud'ko}, \citenamefont {Torikachvili}, \citenamefont {Ni}, \citenamefont
  {Nandi}, \citenamefont {Le\~ao}, \citenamefont {Poulton}, \citenamefont
  {Argyriou}, \citenamefont {Harmon}, \citenamefont {McQueeney}, \citenamefont
  {Canfield},\ and\ \citenamefont {Goldman}}]{Kreyssig2008}%
  \BibitemOpen
  \bibfield  {author} {\bibinfo {author} {\bibfnamefont {A.}~\bibnamefont
  {Kreyssig}}, \bibinfo {author} {\bibfnamefont {M.~A.}\ \bibnamefont {Green}},
  \bibinfo {author} {\bibfnamefont {Y.}~\bibnamefont {Lee}}, \bibinfo {author}
  {\bibfnamefont {G.~D.}\ \bibnamefont {Samolyuk}}, \bibinfo {author}
  {\bibfnamefont {P.}~\bibnamefont {Zajdel}}, \bibinfo {author} {\bibfnamefont
  {J.~W.}\ \bibnamefont {Lynn}}, \bibinfo {author} {\bibfnamefont {S.~L.}\
  \bibnamefont {Bud'ko}}, \bibinfo {author} {\bibfnamefont {M.~S.}\
  \bibnamefont {Torikachvili}}, \bibinfo {author} {\bibfnamefont
  {N.}~\bibnamefont {Ni}}, \bibinfo {author} {\bibfnamefont {S.}~\bibnamefont
  {Nandi}}, \bibinfo {author} {\bibfnamefont {J.~B.}\ \bibnamefont {Le\~ao}},
  \bibinfo {author} {\bibfnamefont {S.~J.}\ \bibnamefont {Poulton}}, \bibinfo
  {author} {\bibfnamefont {D.~N.}\ \bibnamefont {Argyriou}}, \bibinfo {author}
  {\bibfnamefont {B.~N.}\ \bibnamefont {Harmon}}, \bibinfo {author}
  {\bibfnamefont {R.~J.}\ \bibnamefont {McQueeney}}, \bibinfo {author}
  {\bibfnamefont {P.~C.}\ \bibnamefont {Canfield}}, \ and\ \bibinfo {author}
  {\bibfnamefont {A.~I.}\ \bibnamefont {Goldman}},\ }\href {\doibase
  10.1103/PhysRevB.78.184517} {\bibfield  {journal} {\bibinfo  {journal} {Phys.
  Rev. B}\ }\textbf {\bibinfo {volume} {78}},\ \bibinfo {pages} {184517}
  (\bibinfo {year} {2008})}\BibitemShut {NoStop}%
\bibitem [{\citenamefont {Goldman}\ \emph {et~al.}(2009)\citenamefont
  {Goldman}, \citenamefont {Kreyssig}, \citenamefont
  {Proke\ifmmode~\check{s}\else \v{s}\fi{}}, \citenamefont {Pratt},
  \citenamefont {Argyriou}, \citenamefont {Lynn}, \citenamefont {Nandi},
  \citenamefont {Kimber}, \citenamefont {Chen}, \citenamefont {Lee},
  \citenamefont {Samolyuk}, \citenamefont {Le\~ao}, \citenamefont {Poulton},
  \citenamefont {Bud'ko}, \citenamefont {Ni}, \citenamefont {Canfield},
  \citenamefont {Harmon},\ and\ \citenamefont {McQueeney}}]{Goldman2009}%
  \BibitemOpen
  \bibfield  {author} {\bibinfo {author} {\bibfnamefont {A.~I.}\ \bibnamefont
  {Goldman}}, \bibinfo {author} {\bibfnamefont {A.}~\bibnamefont {Kreyssig}},
  \bibinfo {author} {\bibfnamefont {K.}~\bibnamefont
  {Proke\ifmmode~\check{s}\else \v{s}\fi{}}}, \bibinfo {author} {\bibfnamefont
  {D.~K.}\ \bibnamefont {Pratt}}, \bibinfo {author} {\bibfnamefont {D.~N.}\
  \bibnamefont {Argyriou}}, \bibinfo {author} {\bibfnamefont {J.~W.}\
  \bibnamefont {Lynn}}, \bibinfo {author} {\bibfnamefont {S.}~\bibnamefont
  {Nandi}}, \bibinfo {author} {\bibfnamefont {S.~A.~J.}\ \bibnamefont
  {Kimber}}, \bibinfo {author} {\bibfnamefont {Y.}~\bibnamefont {Chen}},
  \bibinfo {author} {\bibfnamefont {Y.~B.}\ \bibnamefont {Lee}}, \bibinfo
  {author} {\bibfnamefont {G.}~\bibnamefont {Samolyuk}}, \bibinfo {author}
  {\bibfnamefont {J.~B.}\ \bibnamefont {Le\~ao}}, \bibinfo {author}
  {\bibfnamefont {S.~J.}\ \bibnamefont {Poulton}}, \bibinfo {author}
  {\bibfnamefont {S.~L.}\ \bibnamefont {Bud'ko}}, \bibinfo {author}
  {\bibfnamefont {N.}~\bibnamefont {Ni}}, \bibinfo {author} {\bibfnamefont
  {P.~C.}\ \bibnamefont {Canfield}}, \bibinfo {author} {\bibfnamefont {B.~N.}\
  \bibnamefont {Harmon}}, \ and\ \bibinfo {author} {\bibfnamefont {R.~J.}\
  \bibnamefont {McQueeney}},\ }\href {\doibase 10.1103/PhysRevB.79.024513}
  {\bibfield  {journal} {\bibinfo  {journal} {Phys. Rev. B}\ }\textbf {\bibinfo
  {volume} {79}},\ \bibinfo {pages} {024513} (\bibinfo {year}
  {2009})}\BibitemShut {NoStop}%
\bibitem [{\citenamefont {Saha}\ \emph {et~al.}(2012)\citenamefont {Saha},
  \citenamefont {Butch}, \citenamefont {Drye}, \citenamefont {Magill},
  \citenamefont {Ziemak}, \citenamefont {Kirshenbaum}, \citenamefont {Zavalij},
  \citenamefont {Lynn},\ and\ \citenamefont {Paglione}}]{Saha2012}%
  \BibitemOpen
  \bibfield  {author} {\bibinfo {author} {\bibfnamefont {S.~R.}\ \bibnamefont
  {Saha}}, \bibinfo {author} {\bibfnamefont {N.~P.}\ \bibnamefont {Butch}},
  \bibinfo {author} {\bibfnamefont {T.}~\bibnamefont {Drye}}, \bibinfo {author}
  {\bibfnamefont {J.}~\bibnamefont {Magill}}, \bibinfo {author} {\bibfnamefont
  {S.}~\bibnamefont {Ziemak}}, \bibinfo {author} {\bibfnamefont
  {K.}~\bibnamefont {Kirshenbaum}}, \bibinfo {author} {\bibfnamefont {P.~Y.}\
  \bibnamefont {Zavalij}}, \bibinfo {author} {\bibfnamefont {J.~W.}\
  \bibnamefont {Lynn}}, \ and\ \bibinfo {author} {\bibfnamefont
  {J.}~\bibnamefont {Paglione}},\ }\href {\doibase 10.1103/PhysRevB.85.024525}
  {\bibfield  {journal} {\bibinfo  {journal} {Phys. Rev. B}\ }\textbf {\bibinfo
  {volume} {85}},\ \bibinfo {pages} {024525} (\bibinfo {year}
  {2012})}\BibitemShut {NoStop}%
\bibitem [{\citenamefont {Ni}\ \emph {et~al.}(2008)\citenamefont {Ni},
  \citenamefont {Nandi}, \citenamefont {Kreyssig}, \citenamefont {Goldman},
  \citenamefont {Mun}, \citenamefont {Bud'ko},\ and\ \citenamefont
  {Canfield}}]{Ni2008}%
  \BibitemOpen
  \bibfield  {author} {\bibinfo {author} {\bibfnamefont {N.}~\bibnamefont
  {Ni}}, \bibinfo {author} {\bibfnamefont {S.}~\bibnamefont {Nandi}}, \bibinfo
  {author} {\bibfnamefont {A.}~\bibnamefont {Kreyssig}}, \bibinfo {author}
  {\bibfnamefont {A.~I.}\ \bibnamefont {Goldman}}, \bibinfo {author}
  {\bibfnamefont {E.~D.}\ \bibnamefont {Mun}}, \bibinfo {author} {\bibfnamefont
  {S.~L.}\ \bibnamefont {Bud'ko}}, \ and\ \bibinfo {author} {\bibfnamefont
  {P.~C.}\ \bibnamefont {Canfield}},\ }\href {\doibase
  10.1103/PhysRevB.78.014523} {\bibfield  {journal} {\bibinfo  {journal} {Phys.
  Rev. B}\ }\textbf {\bibinfo {volume} {78}},\ \bibinfo {pages} {014523}
  (\bibinfo {year} {2008})}\BibitemShut {NoStop}%
\bibitem [{\citenamefont {Cheng}\ \emph {et~al.}(2012)\citenamefont {Cheng},
  \citenamefont {Hu}, \citenamefont {Yuan}, \citenamefont {Dong}, \citenamefont
  {Fang}, \citenamefont {Chen}, \citenamefont {Xu}, \citenamefont {Shi},
  \citenamefont {Zheng}, \citenamefont {Luo},\ and\ \citenamefont
  {Wang}}]{Cheng2012}%
  \BibitemOpen
  \bibfield  {author} {\bibinfo {author} {\bibfnamefont {B.}~\bibnamefont
  {Cheng}}, \bibinfo {author} {\bibfnamefont {B.~F.}\ \bibnamefont {Hu}},
  \bibinfo {author} {\bibfnamefont {R.~H.}\ \bibnamefont {Yuan}}, \bibinfo
  {author} {\bibfnamefont {T.}~\bibnamefont {Dong}}, \bibinfo {author}
  {\bibfnamefont {A.~F.}\ \bibnamefont {Fang}}, \bibinfo {author}
  {\bibfnamefont {Z.~G.}\ \bibnamefont {Chen}}, \bibinfo {author}
  {\bibfnamefont {G.}~\bibnamefont {Xu}}, \bibinfo {author} {\bibfnamefont
  {Y.~G.}\ \bibnamefont {Shi}}, \bibinfo {author} {\bibfnamefont
  {P.}~\bibnamefont {Zheng}}, \bibinfo {author} {\bibfnamefont {J.~L.}\
  \bibnamefont {Luo}}, \ and\ \bibinfo {author} {\bibfnamefont {N.~L.}\
  \bibnamefont {Wang}},\ }\href {\doibase 10.1103/PhysRevB.85.144426}
  {\bibfield  {journal} {\bibinfo  {journal} {Phys. Rev. B}\ }\textbf {\bibinfo
  {volume} {85}},\ \bibinfo {pages} {144426} (\bibinfo {year}
  {2012})}\BibitemShut {NoStop}%
\bibitem [{\citenamefont {Ying}\ \emph {et~al.}(2012)\citenamefont {Ying},
  \citenamefont {Yan}, \citenamefont {Wang}, \citenamefont {Xiang},
  \citenamefont {Cheng}, \citenamefont {Ye},\ and\ \citenamefont
  {Chen}}]{Ying2012}%
  \BibitemOpen
  \bibfield  {author} {\bibinfo {author} {\bibfnamefont {J.~J.}\ \bibnamefont
  {Ying}}, \bibinfo {author} {\bibfnamefont {Y.~J.}\ \bibnamefont {Yan}},
  \bibinfo {author} {\bibfnamefont {A.~F.}\ \bibnamefont {Wang}}, \bibinfo
  {author} {\bibfnamefont {Z.~J.}\ \bibnamefont {Xiang}}, \bibinfo {author}
  {\bibfnamefont {P.}~\bibnamefont {Cheng}}, \bibinfo {author} {\bibfnamefont
  {G.~J.}\ \bibnamefont {Ye}}, \ and\ \bibinfo {author} {\bibfnamefont {X.~H.}\
  \bibnamefont {Chen}},\ }\href {\doibase 10.1103/PhysRevB.85.214414}
  {\bibfield  {journal} {\bibinfo  {journal} {Phys. Rev. B}\ }\textbf {\bibinfo
  {volume} {85}},\ \bibinfo {pages} {214414} (\bibinfo {year}
  {2012})}\BibitemShut {NoStop}%
\bibitem [{\citenamefont {Quirinale}\ \emph {et~al.}(2013)\citenamefont
  {Quirinale}, \citenamefont {Anand}, \citenamefont {Kim}, \citenamefont
  {Pandey}, \citenamefont {Huq}, \citenamefont {Stephens}, \citenamefont
  {Heitmann}, \citenamefont {Kreyssig}, \citenamefont {McQueeney},
  \citenamefont {Johnston},\ and\ \citenamefont {Goldman}}]{Quirinale2013}%
  \BibitemOpen
  \bibfield  {author} {\bibinfo {author} {\bibfnamefont {D.~G.}\ \bibnamefont
  {Quirinale}}, \bibinfo {author} {\bibfnamefont {V.~K.}\ \bibnamefont
  {Anand}}, \bibinfo {author} {\bibfnamefont {M.~G.}\ \bibnamefont {Kim}},
  \bibinfo {author} {\bibfnamefont {A.}~\bibnamefont {Pandey}}, \bibinfo
  {author} {\bibfnamefont {A.}~\bibnamefont {Huq}}, \bibinfo {author}
  {\bibfnamefont {P.~W.}\ \bibnamefont {Stephens}}, \bibinfo {author}
  {\bibfnamefont {T.~W.}\ \bibnamefont {Heitmann}}, \bibinfo {author}
  {\bibfnamefont {A.}~\bibnamefont {Kreyssig}}, \bibinfo {author}
  {\bibfnamefont {R.~J.}\ \bibnamefont {McQueeney}}, \bibinfo {author}
  {\bibfnamefont {D.~C.}\ \bibnamefont {Johnston}}, \ and\ \bibinfo {author}
  {\bibfnamefont {A.~I.}\ \bibnamefont {Goldman}},\ }\href {\doibase
  10.1103/PhysRevB.88.174420} {\bibfield  {journal} {\bibinfo  {journal} {Phys.
  Rev. B}\ }\textbf {\bibinfo {volume} {88}},\ \bibinfo {pages} {174420}
  (\bibinfo {year} {2013})}\BibitemShut {NoStop}%
\bibitem [{\citenamefont {Anand}\ \emph {et~al.}(2014)\citenamefont {Anand},
  \citenamefont {Dhaka}, \citenamefont {Lee}, \citenamefont {Harmon},
  \citenamefont {Kaminski},\ and\ \citenamefont {Johnston}}]{Anand2014}%
  \BibitemOpen
  \bibfield  {author} {\bibinfo {author} {\bibfnamefont {V.~K.}\ \bibnamefont
  {Anand}}, \bibinfo {author} {\bibfnamefont {R.~S.}\ \bibnamefont {Dhaka}},
  \bibinfo {author} {\bibfnamefont {Y.}~\bibnamefont {Lee}}, \bibinfo {author}
  {\bibfnamefont {B.~N.}\ \bibnamefont {Harmon}}, \bibinfo {author}
  {\bibfnamefont {A.}~\bibnamefont {Kaminski}}, \ and\ \bibinfo {author}
  {\bibfnamefont {D.~C.}\ \bibnamefont {Johnston}},\ }\href {\doibase
  10.1103/PhysRevB.89.214409} {\bibfield  {journal} {\bibinfo  {journal} {Phys.
  Rev. B}\ }\textbf {\bibinfo {volume} {89}},\ \bibinfo {pages} {214409}
  (\bibinfo {year} {2014})}\BibitemShut {NoStop}%
\bibitem [{\citenamefont {Canfield}\ and\ \citenamefont
  {Bud'ko}(1997)}]{Canfield1997}%
  \BibitemOpen
  \bibfield  {author} {\bibinfo {author} {\bibfnamefont {P.}~\bibnamefont
  {Canfield}}\ and\ \bibinfo {author} {\bibfnamefont {S.}~\bibnamefont
  {Bud'ko}},\ }\href {\doibase http://dx.doi.org/10.1016/S0925-8388(97)00374-5}
  {\bibfield  {journal} {\bibinfo  {journal} {J. Alloys Comp.}\ }\textbf
  {\bibinfo {volume} {262-263}},\ \bibinfo {pages} {169 } (\bibinfo {year}
  {1997})}\BibitemShut {NoStop}%
\bibitem [{\citenamefont {Naugle}\ \emph {et~al.}(2008)\citenamefont {Naugle},
  \citenamefont {Belevtsev}, \citenamefont {Rathnayaka}, \citenamefont {Lee},\
  and\ \citenamefont {Yeo}}]{Naugle2008}%
  \BibitemOpen
  \bibfield  {author} {\bibinfo {author} {\bibfnamefont {D.~G.}\ \bibnamefont
  {Naugle}}, \bibinfo {author} {\bibfnamefont {B.~I.}\ \bibnamefont
  {Belevtsev}}, \bibinfo {author} {\bibfnamefont {K.~D.~D.}\ \bibnamefont
  {Rathnayaka}}, \bibinfo {author} {\bibfnamefont {S.-I.}\ \bibnamefont {Lee}},
  \ and\ \bibinfo {author} {\bibfnamefont {S.~M.}\ \bibnamefont {Yeo}},\ }\href
  {\doibase http://dx.doi.org/10.1063/1.2829034} {\bibfield  {journal}
  {\bibinfo  {journal} {J. Appl. Phys.}\ }\textbf {\bibinfo {volume} {103}},\
  \bibinfo {eid} {07B718} (\bibinfo {year} {2008})}\BibitemShut {NoStop}%
\bibitem [{\citenamefont {Wang}\ \emph {et~al.}(2009)\citenamefont {Wang},
  \citenamefont {Wu}, \citenamefont {Wu}, \citenamefont {Chen}, \citenamefont
  {Xie}, \citenamefont {Ying}, \citenamefont {Yan}, \citenamefont {Liu},\ and\
  \citenamefont {Chen}}]{Wang2009}%
  \BibitemOpen
  \bibfield  {author} {\bibinfo {author} {\bibfnamefont {X.~F.}\ \bibnamefont
  {Wang}}, \bibinfo {author} {\bibfnamefont {T.}~\bibnamefont {Wu}}, \bibinfo
  {author} {\bibfnamefont {G.}~\bibnamefont {Wu}}, \bibinfo {author}
  {\bibfnamefont {H.}~\bibnamefont {Chen}}, \bibinfo {author} {\bibfnamefont
  {Y.~L.}\ \bibnamefont {Xie}}, \bibinfo {author} {\bibfnamefont {J.~J.}\
  \bibnamefont {Ying}}, \bibinfo {author} {\bibfnamefont {Y.~J.}\ \bibnamefont
  {Yan}}, \bibinfo {author} {\bibfnamefont {R.~H.}\ \bibnamefont {Liu}}, \ and\
  \bibinfo {author} {\bibfnamefont {X.~H.}\ \bibnamefont {Chen}},\ }\href
  {\doibase 10.1103/PhysRevLett.102.117005} {\bibfield  {journal} {\bibinfo
  {journal} {Phys. Rev. Lett.}\ }\textbf {\bibinfo {volume} {102}},\ \bibinfo
  {pages} {117005} (\bibinfo {year} {2009})}\BibitemShut {NoStop}%
\bibitem [{\citenamefont {Ronning}\ \emph {et~al.}(2008)\citenamefont
  {Ronning}, \citenamefont {Klimczuk}, \citenamefont {Bauer}, \citenamefont
  {Volz},\ and\ \citenamefont {Thompson}}]{Ronning2008}%
  \BibitemOpen
  \bibfield  {author} {\bibinfo {author} {\bibfnamefont {F.}~\bibnamefont
  {Ronning}}, \bibinfo {author} {\bibfnamefont {T.}~\bibnamefont {Klimczuk}},
  \bibinfo {author} {\bibfnamefont {E.~D.}\ \bibnamefont {Bauer}}, \bibinfo
  {author} {\bibfnamefont {H.}~\bibnamefont {Volz}}, \ and\ \bibinfo {author}
  {\bibfnamefont {J.~D.}\ \bibnamefont {Thompson}},\ }\href
  {http://stacks.iop.org/0953-8984/20/i=32/a=322201} {\bibfield  {journal}
  {\bibinfo  {journal} {J. Phys. Condens. Matter}\ }\textbf {\bibinfo {volume}
  {20}},\ \bibinfo {pages} {322201} (\bibinfo {year} {2008})}\BibitemShut
  {NoStop}%
\bibitem [{\citenamefont {Ying}\ \emph {et~al.}(2013)\citenamefont {Ying},
  \citenamefont {Liang}, \citenamefont {Luo}, \citenamefont {Yan},
  \citenamefont {Wang}, \citenamefont {Cheng}, \citenamefont {Ye},
  \citenamefont {Ma},\ and\ \citenamefont {Chen}}]{Ying2013}%
  \BibitemOpen
  \bibfield  {author} {\bibinfo {author} {\bibfnamefont {J.~J.}\ \bibnamefont
  {Ying}}, \bibinfo {author} {\bibfnamefont {J.~C.}\ \bibnamefont {Liang}},
  \bibinfo {author} {\bibfnamefont {X.~G.}\ \bibnamefont {Luo}}, \bibinfo
  {author} {\bibfnamefont {Y.~J.}\ \bibnamefont {Yan}}, \bibinfo {author}
  {\bibfnamefont {A.~F.}\ \bibnamefont {Wang}}, \bibinfo {author}
  {\bibfnamefont {P.}~\bibnamefont {Cheng}}, \bibinfo {author} {\bibfnamefont
  {G.~J.}\ \bibnamefont {Ye}}, \bibinfo {author} {\bibfnamefont {J.~Q.}\
  \bibnamefont {Ma}}, \ and\ \bibinfo {author} {\bibfnamefont {X.~H.}\
  \bibnamefont {Chen}},\ }\href
  {http://stacks.iop.org/0295-5075/104/i=6/a=67005} {\bibfield  {journal}
  {\bibinfo  {journal} {EPL}\ }\textbf {\bibinfo {volume}
  {104}},\ \bibinfo {pages} {67005} (\bibinfo {year} {2013})}\BibitemShut
  {NoStop}%
\bibitem [{\citenamefont {Xiao}\ \emph {et~al.}(2006)\citenamefont {Xiao},
  \citenamefont {Hu}, \citenamefont {Almasan}, \citenamefont {Sayles},\ and\
  \citenamefont {Maple}}]{Xiao2006}%
  \BibitemOpen
  \bibfield  {author} {\bibinfo {author} {\bibfnamefont {H.}~\bibnamefont
  {Xiao}}, \bibinfo {author} {\bibfnamefont {T.}~\bibnamefont {Hu}}, \bibinfo
  {author} {\bibfnamefont {C.~C.}\ \bibnamefont {Almasan}}, \bibinfo {author}
  {\bibfnamefont {T.~A.}\ \bibnamefont {Sayles}}, \ and\ \bibinfo {author}
  {\bibfnamefont {M.~B.}\ \bibnamefont {Maple}},\ }\href {\doibase
  10.1103/PhysRevB.73.184511} {\bibfield  {journal} {\bibinfo  {journal} {Phys.
  Rev. B}\ }\textbf {\bibinfo {volume} {73}},\ \bibinfo {pages} {184511}
  (\bibinfo {year} {2006})}\BibitemShut {NoStop}%
\bibitem [{\citenamefont {Nagamiya}(1951)}]{Nagamiya1951a}%
  \BibitemOpen
  \bibfield  {author} {\bibinfo {author} {\bibfnamefont {T.}~\bibnamefont
  {Nagamiya}},\ }\href {\doibase 10.1143/ptp/6.3.342} {\bibfield  {journal}
  {\bibinfo  {journal} {Prog. Theor. Phys.}\ }\textbf {\bibinfo {volume} {6}},\
  \bibinfo {pages} {342} (\bibinfo {year} {1951})}\BibitemShut {NoStop}%
\bibitem [{\citenamefont {Yosida}(1951)}]{Yosida1951}%
  \BibitemOpen
  \bibfield  {author} {\bibinfo {author} {\bibfnamefont {K.}~\bibnamefont
  {Yosida}},\ }\href {\doibase 10.1143/ptp/6.5.691} {\bibfield  {journal}
  {\bibinfo  {journal} {Prog. Theor. Phys.}\ }\textbf {\bibinfo {volume} {6}},\
  \bibinfo {pages} {691} (\bibinfo {year} {1951})}\BibitemShut {NoStop}%
\bibitem [{\citenamefont {Nagamiya}\ \emph {et~al.}(1955)\citenamefont
  {Nagamiya}, \citenamefont {Yosida},\ and\ \citenamefont
  {Kubo}}]{Nagamiya1955}%
  \BibitemOpen
  \bibfield  {author} {\bibinfo {author} {\bibfnamefont {T.}~\bibnamefont
  {Nagamiya}}, \bibinfo {author} {\bibfnamefont {K.}~\bibnamefont {Yosida}}, \
  and\ \bibinfo {author} {\bibfnamefont {R.}~\bibnamefont {Kubo}},\ }\href
  {\doibase 10.1080/00018735500101154} {\bibfield  {journal} {\bibinfo
  {journal} {Adv. Phys.}\ }\textbf {\bibinfo {volume} {4}},\ \bibinfo {pages}
  {1} (\bibinfo {year} {1955})}\BibitemShut {NoStop}%
\bibitem [{\citenamefont {Weyeneth}\ \emph {et~al.}(2011)\citenamefont
  {Weyeneth}, \citenamefont {Moll}, \citenamefont {Puzniak}, \citenamefont
  {Ninios}, \citenamefont {Balakirev}, \citenamefont {McDonald}, \citenamefont
  {Chan}, \citenamefont {Zhigadlo}, \citenamefont {Katrych}, \citenamefont
  {Bukowski}, \citenamefont {Karpinski}, \citenamefont {Keller}, \citenamefont
  {Batlogg},\ and\ \citenamefont {Balicas}}]{Weyeneth2011}%
  \BibitemOpen
  \bibfield  {author} {\bibinfo {author} {\bibfnamefont {S.}~\bibnamefont
  {Weyeneth}}, \bibinfo {author} {\bibfnamefont {P.~J.~W.}\ \bibnamefont
  {Moll}}, \bibinfo {author} {\bibfnamefont {R.}~\bibnamefont {Puzniak}},
  \bibinfo {author} {\bibfnamefont {K.}~\bibnamefont {Ninios}}, \bibinfo
  {author} {\bibfnamefont {F.~F.}\ \bibnamefont {Balakirev}}, \bibinfo {author}
  {\bibfnamefont {R.~D.}\ \bibnamefont {McDonald}}, \bibinfo {author}
  {\bibfnamefont {H.~B.}\ \bibnamefont {Chan}}, \bibinfo {author}
  {\bibfnamefont {N.~D.}\ \bibnamefont {Zhigadlo}}, \bibinfo {author}
  {\bibfnamefont {S.}~\bibnamefont {Katrych}}, \bibinfo {author} {\bibfnamefont
  {Z.}~\bibnamefont {Bukowski}}, \bibinfo {author} {\bibfnamefont
  {J.}~\bibnamefont {Karpinski}}, \bibinfo {author} {\bibfnamefont
  {H.}~\bibnamefont {Keller}}, \bibinfo {author} {\bibfnamefont
  {B.}~\bibnamefont {Batlogg}}, \ and\ \bibinfo {author} {\bibfnamefont
  {L.}~\bibnamefont {Balicas}},\ }\href {\doibase 10.1103/PhysRevB.83.134503}
  {\bibfield  {journal} {\bibinfo  {journal} {Phys. Rev. B}\ }\textbf {\bibinfo
  {volume} {83}},\ \bibinfo {pages} {134503} (\bibinfo {year}
  {2011})}\BibitemShut {NoStop}%
\bibitem [{\citenamefont {Watson}\ \emph {et~al.}(2014)\citenamefont {Watson},
  \citenamefont {McCollam}, \citenamefont {Blake}, \citenamefont {Vignolles},
  \citenamefont {Drigo}, \citenamefont {Mazin}, \citenamefont {Guterding},
  \citenamefont {Jeschke}, \citenamefont {Valenti}, \citenamefont {Ni},
  \citenamefont {Cava},\ and\ \citenamefont {Coldea}}]{Watson2014}%
  \BibitemOpen
  \bibfield  {author} {\bibinfo {author} {\bibfnamefont {M.~D.}\ \bibnamefont
  {Watson}}, \bibinfo {author} {\bibfnamefont {A.}~\bibnamefont {McCollam}},
  \bibinfo {author} {\bibfnamefont {S.~F.}\ \bibnamefont {Blake}}, \bibinfo
  {author} {\bibfnamefont {D.}~\bibnamefont {Vignolles}}, \bibinfo {author}
  {\bibfnamefont {L.}~\bibnamefont {Drigo}}, \bibinfo {author} {\bibfnamefont
  {I.~I.}\ \bibnamefont {Mazin}}, \bibinfo {author} {\bibfnamefont
  {D.}~\bibnamefont {Guterding}}, \bibinfo {author} {\bibfnamefont {H.~O.}\
  \bibnamefont {Jeschke}}, \bibinfo {author} {\bibfnamefont {R.}~\bibnamefont
  {Valenti}}, \bibinfo {author} {\bibfnamefont {N.}~\bibnamefont {Ni}},
  \bibinfo {author} {\bibfnamefont {R.}~\bibnamefont {Cava}}, \ and\ \bibinfo
  {author} {\bibfnamefont {A.~I.}\ \bibnamefont {Coldea}},\ }\href {\doibase
  10.1103/PhysRevB.89.205136} {\bibfield  {journal} {\bibinfo  {journal} {Phys.
  Rev. B}\ }\textbf {\bibinfo {volume} {89}},\ \bibinfo {pages} {205136}
  (\bibinfo {year} {2014})}\BibitemShut {NoStop}%
\bibitem [{\citenamefont {Xu}\ \emph {et~al.}(2010)\citenamefont {Xu},
  \citenamefont {Carrington}, \citenamefont {Coldea}, \citenamefont
  {Enayati-Rad}, \citenamefont {Narduzzo}, \citenamefont {Horii},\ and\
  \citenamefont {Hussey}}]{Xu2010}%
  \BibitemOpen
  \bibfield  {author} {\bibinfo {author} {\bibfnamefont {X.}~\bibnamefont
  {Xu}}, \bibinfo {author} {\bibfnamefont {A.}~\bibnamefont {Carrington}},
  \bibinfo {author} {\bibfnamefont {A.~I.}\ \bibnamefont {Coldea}}, \bibinfo
  {author} {\bibfnamefont {A.}~\bibnamefont {Enayati-Rad}}, \bibinfo {author}
  {\bibfnamefont {A.}~\bibnamefont {Narduzzo}}, \bibinfo {author}
  {\bibfnamefont {S.}~\bibnamefont {Horii}}, \ and\ \bibinfo {author}
  {\bibfnamefont {N.~E.}\ \bibnamefont {Hussey}},\ }\href {\doibase
  10.1103/PhysRevB.81.224435} {\bibfield  {journal} {\bibinfo  {journal} {Phys.
  Rev. B}\ }\textbf {\bibinfo {volume} {81}},\ \bibinfo {pages} {224435}
  (\bibinfo {year} {2010})}\BibitemShut {NoStop}%
\bibitem [{\citenamefont {Waldmann}\ \emph {et~al.}(2002)\citenamefont
  {Waldmann}, \citenamefont {Zhao},\ and\ \citenamefont
  {Thompson}}]{Waldmann2002}%
  \BibitemOpen
  \bibfield  {author} {\bibinfo {author} {\bibfnamefont {O.}~\bibnamefont
  {Waldmann}}, \bibinfo {author} {\bibfnamefont {L.}~\bibnamefont {Zhao}}, \
  and\ \bibinfo {author} {\bibfnamefont {L.~K.}\ \bibnamefont {Thompson}},\
  }\href {\doibase 10.1103/PhysRevLett.88.066401} {\bibfield  {journal}
  {\bibinfo  {journal} {Phys. Rev. Lett.}\ }\textbf {\bibinfo {volume} {88}},\
  \bibinfo {pages} {066401} (\bibinfo {year} {2002})}\BibitemShut {NoStop}%
\bibitem [{\citenamefont {Tajima}\ \emph {et~al.}(2008)\citenamefont {Tajima},
  \citenamefont {Yoshida}, \citenamefont {Matsuda}, \citenamefont {Nara},
  \citenamefont {Kajita}, \citenamefont {Nishio}, \citenamefont {Hanasaki},
  \citenamefont {Naito},\ and\ \citenamefont {Inabe}}]{Tajima2008}%
  \BibitemOpen
  \bibfield  {author} {\bibinfo {author} {\bibfnamefont {H.}~\bibnamefont
  {Tajima}}, \bibinfo {author} {\bibfnamefont {G.}~\bibnamefont {Yoshida}},
  \bibinfo {author} {\bibfnamefont {M.}~\bibnamefont {Matsuda}}, \bibinfo
  {author} {\bibfnamefont {K.}~\bibnamefont {Nara}}, \bibinfo {author}
  {\bibfnamefont {K.}~\bibnamefont {Kajita}}, \bibinfo {author} {\bibfnamefont
  {Y.}~\bibnamefont {Nishio}}, \bibinfo {author} {\bibfnamefont
  {N.}~\bibnamefont {Hanasaki}}, \bibinfo {author} {\bibfnamefont
  {T.}~\bibnamefont {Naito}}, \ and\ \bibinfo {author} {\bibfnamefont
  {T.}~\bibnamefont {Inabe}},\ }\href {\doibase 10.1103/PhysRevB.78.064424}
  {\bibfield  {journal} {\bibinfo  {journal} {Phys. Rev. B}\ }\textbf {\bibinfo
  {volume} {78}},\ \bibinfo {pages} {064424} (\bibinfo {year}
  {2008})}\BibitemShut {NoStop}%
\bibitem [{\citenamefont {Torizuka}\ \emph {et~al.}(2013)\citenamefont
  {Torizuka}, \citenamefont {Tajima}, \citenamefont {Inoue}, \citenamefont
  {Hanasaki}, \citenamefont {Matsuda}, \citenamefont {E.~C.~Yu}, \citenamefont
  {Naito},\ and\ \citenamefont {Inabe}}]{Torizuka2013}%
  \BibitemOpen
  \bibfield  {author} {\bibinfo {author} {\bibfnamefont {K.}~\bibnamefont
  {Torizuka}}, \bibinfo {author} {\bibfnamefont {H.}~\bibnamefont {Tajima}},
  \bibinfo {author} {\bibfnamefont {M.}~\bibnamefont {Inoue}}, \bibinfo
  {author} {\bibfnamefont {N.}~\bibnamefont {Hanasaki}}, \bibinfo {author}
  {\bibfnamefont {M.}~\bibnamefont {Matsuda}}, \bibinfo {author} {\bibfnamefont
  {D.}~\bibnamefont {E.~C.~Yu}}, \bibinfo {author} {\bibfnamefont
  {T.}~\bibnamefont {Naito}}, \ and\ \bibinfo {author} {\bibfnamefont
  {T.}~\bibnamefont {Inabe}},\ }\href {\doibase 10.7566/JPSJ.82.034719}
  {\bibfield  {journal} {\bibinfo  {journal} {J. Phys. Soc. Jpn.}\ }\textbf
  {\bibinfo {volume} {82}},\ \bibinfo {pages} {034719} (\bibinfo {year}
  {2013})}\BibitemShut {NoStop}%
\end{thebibliography}

%

\end{document}